\documentclass[aps,pra,twocolumn,showpacs,superscriptaddress,groupedaddress]{revtex4-1}  % for review and submission
\usepackage{amssymb,amsmath,graphicx}
\usepackage{graphicx,wrapfig,lipsum}
\usepackage[usenames,dvipsnames]{xcolor}
\usepackage{graphicx}  % needed for figures
\usepackage{bm}        % for math
\usepackage{amssymb}   % for math
\usepackage{braket}
\usepackage{mathtools}
\usepackage{booktabs} %for table
\usepackage{commath}
\usepackage[dvipsnames]{xcolor}
\usepackage{hyperref}
\usepackage{mathtools}
\newcommand{\vect}[1]{\mathbf{#1}}

\newcommand{\bsf}[1]{\textsf{\textbf{#1}}}
\usepackage{appendix}
\usepackage{subfigure}
\usepackage{ulem}
\newcommand{\grad}[1]{\vect{\nabla}#1}
\renewcommand{\t}[1]{\widetilde{#1}}
\DeclareMathOperator{\Tr}{Tr}

\begin{document}
\title{Active Caustics}
\author{Rahul Chajwa $^{1,2}$, Rajarshi $^2$, Rama Govindarajan $^2$, Sriram Ramaswamy $^{3,2}$}
\affiliation{$^1$Department of Bioengineering, Stanford University, Stanford, CA 94305 USA. \\ 
$^2$International Centre for Theoretical Sciences, Tata Institute of Fundamental Research, Bengaluru 560 089 \\ 
$^3$Centre for Condensed Matter Theory, Department of Physics, Indian Institute of Science, Bengaluru 560 012.}

%\date{\today}
%\maketitle
\begin{abstract}
Inertial particles (IPs) in vortical fluid flow cluster strongly, forming singular structures termed \textit{caustics} for their resemblance to focal surfaces in optics. Here we show that such extreme aggregation onto low-dimensional submanifolds can arise \textit{without} mechanical inertia for self-propelled particles (SPPs), through a formal correspondence between the dynamics of IPs and SPPs in a generic background flow. We establish that a singular perturbation underlies caustics formation by SPPs around a single vortex, and numerical studies of SPPs in two-dimensional Navier-Stokes turbulence reveal intense caustics in straining regions of the flow, peaking at intermediate levels of self-propulsion. Our work offers a route to singularly high local concentrations in a macroscopically dilute suspension of zero-Reynolds-number swimmers. Caustics generate burst-like encounters through 
%sudden 
large relative velocities %\rgc{what are sudden relative velocities} 
%between nearby particles, 
between neighboring swimmers, with potentially significant implications for communication and sexual reproduction. An intriguing open direction is whether the active turbulence of a suspension of swimming microbes could serve to generate caustics in its own concentration.

\end{abstract}
\maketitle 

\section{Introduction}
In geometrical optics, the boundary of a family of focused light rays is termed a \textit{caustic}. The extremal character of such a limiting surface or curve (in three or two dimensions, respectively) implies a divergent intensity \cite{Berry1977} By precise analogy the term caustic is applied to singularities in particle concentration arising from the crossing of trajectories -- of dark matter on cosmic scales \cite{Feldbrugge2018} or of solute in suspensions \cite{Wilkinson_2005}, the latter being the focus here. The growth of raindrops in clouds \cite{PINSKY19971177}, the settling of atmospheric pollutants~\cite{Fernando2010}, the strength of the biological pump in the oceans~\cite{Boyd}, and a wide range of industrial processes \cite{SAMBORSKA2022110960}, hinge on collisions and coalescence between particles or droplets suspended in background flows. When particles have appreciable inertia, the coupling between their velocity and the local flow profile can lead to inhomogeneities in their spatial distribution \cite{Lillo2004}. Heavy particles in a turbulent flow are known to get centrifuged out of regions of high vorticity, displaying finite-time singularities in particle trajectories that amplify collision rates, aggregation, and caustics~\cite{MR1983, Maxey1986, Croor2015, Wilkinson_2005}, offering a mechanism for droplet growth in clouds~\cite{falkovich2002, Croor2017}, for example. In this article we explore theoretically the possibility of such singular enhancement of encounters in suspensions of motile particles \cite{lauga2009hydrodynamics} {\it without} particle inertia. 

Swimmers in vortical and turbulent flows, through the coupling of their orientation to velocity gradients, present rich dynamical behaviors: accumulation at vortex boundaries \cite{Torney2007}, gyrotactic patchiness in vortical and turbulent flows \cite{Durham2011, Durham2013, Zhan2014}, crossing of transport barriers in chaotic flows \cite{Khurana2011}, and preferential sampling of downwelling regions by marine microorganisms \cite{pedley1992hydrodynamic, Durham2013, PhysRevLett.116.108104}. Micro-swimmers in externally driven flows display focusing \cite{Kessler1985}, aggregation \cite{GENIN20043, Ardekani2012}, and expulsion out of vortical regions \cite{Sokolov2016}, in a manner reminiscent of inertial particles \cite{Croor2015, Wilkinson_2005}. These works established that motility and shape when coupled to background vortical flows generate clustering and transport phenomena absent for passive tracers. Durham et al. \cite{Durham2013}, for example, remark on the visual similarity of fractal clustering of active particles in flow and cloud formation by droplets \cite{falkovich2002}. Here, we establish a more formal correspondence by identifying caustics as a unifying dynamical mechanism.

% \begin{figure}[t]
%  \begin{center}
%  \includegraphics[width=7.5 cm]{mainFigures/Fig1_reduced.jpeg}
%  %\hspace{0cm}\newline
%  \caption{\label{Fig1} \textbf{Active dimer model for self-propelled particles in ambient flow} (a) Schematic of an active dimer of extension $\vect{w}$ in a flow $\vect{U}$. (b) Caustics based on the inner solution are marked by the intersection of representative trajectories (blue circles) of particles starting at closely separated initial radial distances, with $\alpha = 0.1$. A continuous variation in $\tilde{r}_0$ would give a continuous curve. The inset (Photo credit: R. Chajwa) is an image of optical caustics with a similar cusp on the surface of coffee in a mug.} %\remarkSR{Check source for coffee photo or take a photo yourself.}}  -- This is my own coffee mug :) 
%  \label{Figure1}
%  \end{center}
%  \end{figure}

\begin{figure*}[t]
 \begin{center}
 \includegraphics[width=16.5 cm]{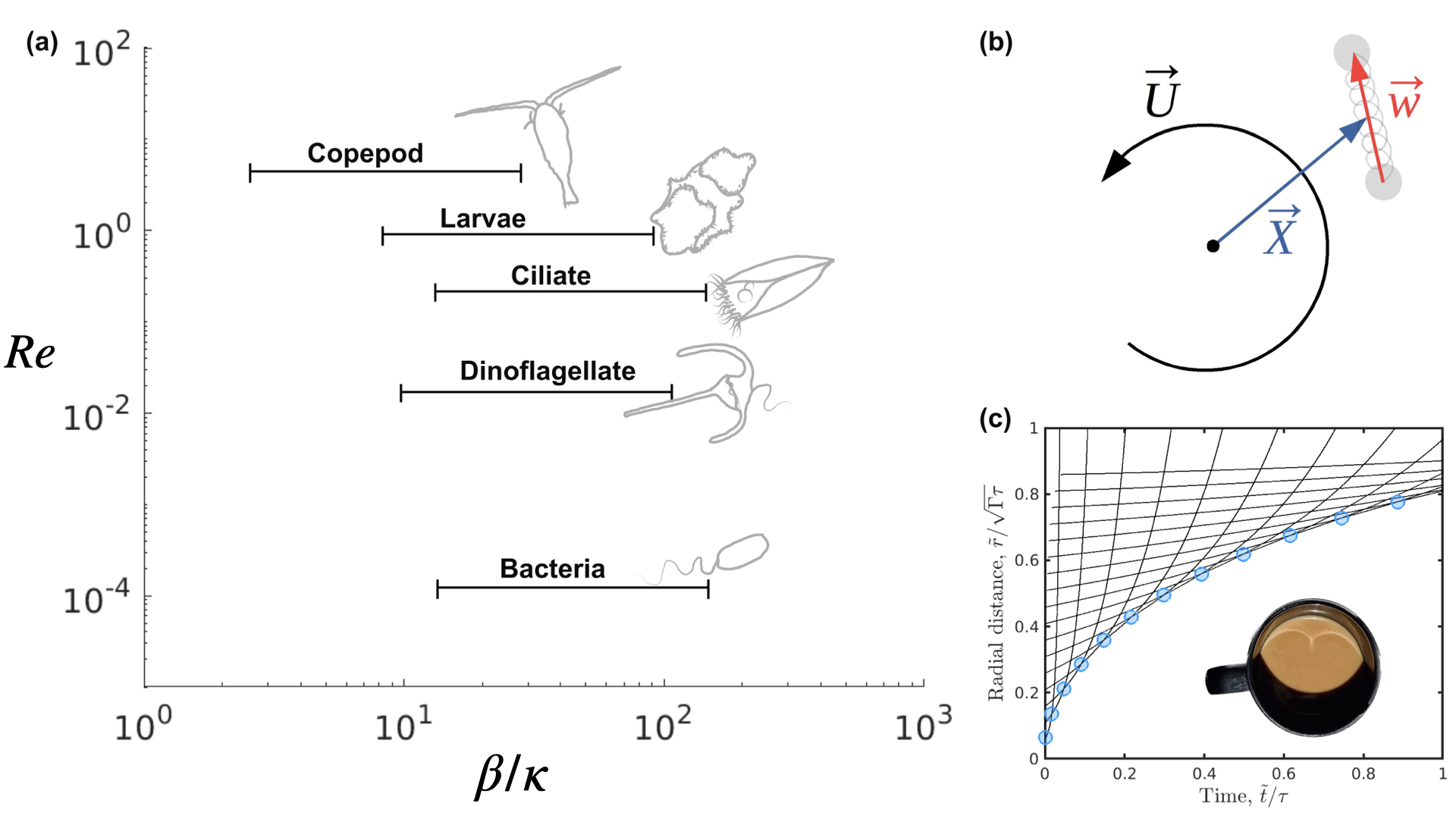}
 %\hspace{0cm}\newline
 \caption{\label{Fig1} \textbf{Active dimer model for self-propelled particles in ambient flow} (a) We depict typical values of the ratio of flow time scale $1/\norm{\nabla \vect{U}}$ to swimming time scale $1/\beta$, and Reynolds number $Re$ for a marine bacterium \textit{Vibrio alginolyticus} \cite{%10.7554/eLife.22140, 
 doi:10.1073/pnas.1602307113}, various dinoflagellates, ciliates \cite{Lauga2019}, invertebrate larvae \cite{FUCHS2016109} and copepods \cite{FUCHS2016109}. For $\norm{\nabla \vect{U}}$ we substitute the Kolmogorov shear rate for the range of previously measured energy dissipation rates in the upper mixed layer of the ocean $10^{-8} - 10^{-6}$ m$^{2}$s$^{-3}$ \cite{StockerARFM2012} [see Appendix \ref{plankton}]. (b) Schematic of an active dimer of extension $\vect{w}$ in a flow $\vect{U}$. (c) Caustics based on the inner solution are marked by the intersection of representative trajectories (blue circles) of particles starting at closely separated initial radial distances, with $\alpha = 0.1$. A continuous variation in $\tilde{r}_0$ would give a continuous curve. The inset (Photo credit: R. Chajwa) is an image of optical caustics with a similar cusp on the surface of coffee in a mug.} %\remarkSR{Check source for coffee photo or take a photo yourself.}}  -- This is my own coffee mug :) 
 \label{Figure1}
 \end{center}
 \end{figure*}

Biological motility in a fluid medium \cite{Tuval2020} is dynamic, and flows can strongly influence the dispersal of swimmers \cite{PhysRevLett.129.064502,  PhysRevFluids.6.L012501}. The competition between autonomous motion and systematic reorientation by the ambient flow field $\vect{U}$ is characterized by the local ratio $\beta/\norm{\nabla \vect{U}}$ where $1/\beta$ is the time it takes for a swimmer to move its own body length. When $\vect{U}$ is unsteady and/or spatially heterogeneous, this ratio is dynamic along a swimmer trajectory, and thus it is convenient to define an average quantity $\beta/\kappa$, where $\kappa \equiv \sqrt{\langle \nabla \vect{U : \nabla\vect{U}} \rangle}$ is the root mean square value. %\sr{We will define the average $\langle ... \rangle$ more precisely later in the article.} \remarkSR{[Rahul, where is the average defined?]} 
Fig. \ref{Figure1} (a) depicts typical values of $\beta/\kappa$ and Reynolds ($Re$) numbers for various swimmers, based on published data on swimming speed and size, and typical shear-rates in the upper mixed layer of the open ocean \cite{StockerARFM2012} [see Appendix \ref{plankton}]. In complex natural environments, such as oceans, neither particle inertia nor activity can be directly controlled, and `collisional' aggregation \cite{pumir2016} may arise from both inertia and activity [hence $Re$ as a separate axis in Fig. \ref{Figure1} (a)]. We focus on swimmers for which $Re$ and Stokes numbers $St$ are vanishingly small, with self-propelling stresses small compared to those created by the ambient flow. We therefore ignore particle inertia as well as the effect of the particles on the flow, neglecting interparticle hydrodynamic and steric interactions. Although their inertia is negligible, their motion is persistent because of self-propulsion \cite{Durham2013}.

We ask whether inertialess swimmers in flow \cite{pedley1992hydrodynamic} can display caustics analogous to those of inertial particles (IPs) \cite{Wilkinson_2005,Croor2015}, where the calculated world lines of suspended particles cross, rendering the solute velocity field multivalued. We investigate this previously unexplored possibility theoretically, and show that a mathematically divergent particle number-density can emerge  in the minimal setting of a dilute suspension of neutrally buoyant self-propelled particles, \textit{without} particle inertia, in two-dimensional vortical flows. We examine the flow-coupled dynamics of two simple models of single motile particles: Hookean and preferred-length active dimers. In the presence of external noise, these correspond to active Ornstein-Uhlenbeck particles (AOUPs) \cite{Howfar2016, Bonilla2019}, and active Brownian particles (ABPs) \cite{Romanczuk2012}, respectively. Using a singular perturbation analysis, we show that non-inertial active particles, like passive IPs, can display caustics even near a single vortex, a building block of turbulence. We establish how far the analogy with IPs may be carried, and where the two differ fundamentally. We extend these findings to a continuous particle phase of dimer suspensions, which removes interparticle separation as a free parameter and allows an unambiguous demarcation of the regimes of caustic formation.

The dynamics around a single vortex forms the basis for understanding the behavior of active particles in unsteady vortical flows. We study the dynamics of preferred-length active dimers in homogeneous, isotropic turbulence through a pseudospectral direct numerical simulation, and establish caustic formation, with an intensity that peaks at an intermediate value of $\beta/\kappa$ and show bounds on its structural stability with respect to orientational noise. The coupling of motility to ambient flow plays a central role in the lives of microscopic marine plankton \cite{Tuval2020, Krishnamurthy2020, mousavi2023efficient}. By cataloging previously collected data on $\beta/\kappa$ of such microswimmers in marine environments, we predict that they can display caustics in a vortex-laden turbulent ecosystem in the upper ocean. Our study reveals that active caustics can arise generically in vortical flows even in the absence of turbulence, or organism-specific sensory mechanisms like gyrotaxis or chemotaxis \cite{Pedley1992}. Active caustics produces sudden burst-like encounters between nearby particles, offering a new route to increased opportunities for communication and sexual reproduction \cite{BUSKEY199813} among swimmers in marine environments \cite{Tuval2020, StockerARFM2012}.%, conceptually distinct from simple clustering. \remarkSR{[How distinct? Is our case that good?]}

\section{Correspondence between Active and Inertial Particles in a Flow}
The motion of an IP with position vector $\vect{X}$ in a background flow-field $\vect{U}$ is governed by the Maxey-Riley~\cite{MR1983} equation which, to leading order in gradients, reads $\dot{\vect{X}} = \vect{v}$ and $\text{St} \, \dot{\vect{v}}=  \, (\vect{U}-\vect{v})$, when non-dimensionalised using a characteristic particle length scale $d$ and a flow velocity scale $U_0$. The Stokes number $\text{St} = \tau U_0/d$ is a non-dimensional measure of inertia, characterised by the relaxation time $\tau$ ($=$ mass/Stokes drag coefficient) of a particle. The centrifugation of these particles away from the vortex center results in the formation of caustics within a critical distance from the vortex origin \cite{Croor2015}. Setting  $\text{St}=0$ yields tracer particles, which move with $\vect{v}=\vect{U}$.

The Hamiltonian structure of bound orbits of microswimmers in certain flows \cite{Stark2016, Lushi2015, Shelley2019, Arguedas_Leiva_2020} points to an effective inertial character in the dynamics of active Stokesian suspensions under imposed flows, wherein the orientation vector plays a role analogous to momentum \cite{Chajwa2019, Ronojoy2020, Chajwa2020}. Building on this observation, we construct below an effective inertial description for active particles in a generic background flow. The active particles of interest to us are orientable objects with negligible inertia, whose dynamics involves a centroid position $\mathbf{X}$ and an end-to-end vector $\mathbf{w}$ that describes their spatial extension and the direction in which they propel themselves [Fig. \ref{Fig1} (b)]. The case of fixed or preferred length $|\mathbf{w}|$ and self-propulsion at a constant speed along $\hat{\mathbf{w}}$ corresponds to the ABP \cite{Romanczuk2012} mentioned above. However, to demonstrate analytically that caustics and the consequent discontinuities in particle number densities can arise due to activity, even without inertia, we will begin with a simple and analytically tractable dynamical model for $\mathbf{w}$, namely, the Hookean active dimer, placed in an imposed background flow field $\mathbf{U}$. When Gaussian white noise is added to the dynamics of $\mathbf{w}$, the correlations of the relative coordinate of the dimer will decay exponentially in time, and the dynamics of its centroid will be that of an active Ornstein-Uhlenbeck \cite{Howfar2016, Bonilla2019} particle, an instructive and analytically tractable test bed for the development of ideas in active matter.

We work at zero Stokes number and thus neglect particle inertia, but, as we see below, self-propulsion allows the particle to cross streamlines. In the absence of translational diffusion the equations of motion for $\mathbf{X}$ and $\mathbf{w}$ then take the first-order form  
\begin{subequations} 
\label{eqn:dimer}
\begin{equation}
\dot{\vect{X}} = \mu \vect{F} \, + \vect{U} \, + \beta \vect{w} \, \equiv \, \vect{v},  \label{eqn:dimerX}
\end{equation}
\begin{equation}
\dot{\vect{w}} = -\frac{\vect{w}}{\tau} \, + \, ( \alpha \, \bsf{S} + \bsf{A} \,) \cdot \,\vect{w}  - %l_0 
\ell^2 \nabla^{2} \vect{U} \, 
%+  {\sqrt{\frac{2D}{\tau}}} \bm{\eta}, 
+  \sqrt{2D} \bm{\eta}, 
\label{eqn:dimerw}
\end{equation}
\end{subequations}  
where $\mu$ is the Stokesian mobility of the particle, and $\vect{U}$, $\nabla^2\vect{U}$ and the external force field $\vect{F}$ are evaluated at $\vect{X}(t)$.  
In \eqref{eqn:dimerX} $\beta$, with units of inverse time but indefinite sign, endows a dimer with self-propulsion proportional to its extension, parallel or antiparallel to $\vect{w}$ depending on sgn$(\beta)$. The polar flow-alignment parameter $\ell$, with units of length, orients the dimer along a locally parabolic flow, and vanishes for an apolar, i.e., fore-aft symmetric, particle. A term of the $\ell$ type appears in the context of the dynamics of a \textit{collective} orientation vector, in \cite{maitra2014activating}. Mathematically, in a gradient expansion, it is the lowest-order coupling of orientation to flow that is sensitive to the orientation \textit{vector} rather than simply the headless \textit{axis} of the particle. Physically: locally in an unbounded fluid an object characterized by a vector $\vect{w}$ can align not with $\vect{U}$ itself, as that is not a frame-invariant notion, but with a Galilean-invariant vector made from $\vect{U}$, the simplest being $\nabla^2\vect{U}$. For a system confined between planar walls, which provide an absolute frame of reference, a term simply proportional to $\vect{U}$ is permitted, as discussed in \cite{shimoyama1996collective,brotto2013hydrodynamics,kumar2014flocking}.  
For a review of microswimmers in imposed flow-fields, though without the polar coupling $\ell$, see \cite{Stark2016}.  Apolar flow-orientation couplings \cite{Jeffery1922} enter through the strain-rate and vorticity tensors $\bsf{S}\equiv (\nabla \mathbf{U} + \nabla \mathbf{U}^\top)/2$ and $\bsf{A} \equiv (\nabla \mathbf{U} - \nabla \mathbf{U}^\top)/2$ respectively, with a response parameter $\alpha$. While the original Jeffery equations are for a rigid ellipsoid, in this study $\vect{w}$ does not follow a rigidity constraint, relaxing the requirement of a transverse projection operator $(\mathbb{I} - \vect{ww})$ \cite{Jeffery1922} in \eqref{eqn:dimerw} . Since the shape of the dimer depends on $|\vect{w}|$, $\alpha$ in our model does not admit the direct geometric interpretation that underlies the shape-dependent coupling in \cite{Jeffery1922}. 
In \eqref{eqn:dimerw} $\bm{\eta}$ is a zero-mean, isotropic, gaussian white noise with unit variance. With this preamble we proceed to derive an effective inertial equation of an active particle in a background flow.

 For $\beta \neq 0$, \eqref{eqn:dimerX} and \eqref{eqn:dimerw} yield an equation for the total active-particle velocity ${\bf v}$ [see \eqref{eqn:dimerX}] viewed as a dynamical variable
\begin{align*}
\frac{d \mathbf{v}}{dt}  = & \mu \mathbf{v} \cdot \nabla \mathbf{F} + \mathbf{v} \cdot \nabla \mathbf{U} -\frac{\beta}{\tau} \mathbf{w} + \beta \left( \alpha \bsf{S} + \bsf{A}\right) \cdot \mathbf{w} \\ \nonumber & +  \partial_{t}\mathbf{U} - \beta \ell^{2} \nabla ^{2} \mathbf{U} + \beta \sqrt{2D} \mathbf{\eta}
\end{align*}
\begin{align*}
 = & \left[ - \frac{1}{\tau} \bsf{I} + (\alpha \bsf{S} + \bsf{A})\right] \cdot (\mathbf{v} - \mathbf{U} - \mu \mathbf{F}) \\ \nonumber
 & + \left[ \mu (\nabla \mathbf{F})^{\intercal}  + (\nabla \mathbf{U})^{\intercal}\right] \cdot \mathbf{v} + \partial_{t} \mathbf{U} - \beta \ell^{2} \nabla^{2} \mathbf{U} + \beta \sqrt{2D} \mathbf{\eta}.
\end{align*}
Using
 \begin{align*}
    \bsf{A} \cdot \mathbf{v} + (\nabla \mathbf{U})^{\intercal} \cdot \mathbf{v} & = \frac{1}{2} \left[ \nabla \mathbf{U} - (\nabla \mathbf{U})^{\intercal}\right] \cdot \mathbf{v} + (\nabla \mathbf{U})^{\intercal} \cdot \mathbf{v} \\
    & = \frac{1}{2} \left[ \nabla \mathbf{U} + (\nabla \mathbf{U})^{\intercal}\right] \cdot \mathbf{v} \\
    & =  \bsf{S} \cdot \mathbf{v},
 \end{align*}
and multiplying both sides by $\tau/\mu$, gives
\begin{align}
\frac{\tau}{\mu} \frac{d \mathbf{v}}{dt} & = \left[\bsf{I} - \tau (\alpha \bsf{S} + \bsf{A})\right]\cdot (\mathbf{F} + \frac{1}{\mu} \mathbf{U}) + \frac{\tau}{\mu}( \partial_{t} - {\beta \ell^2} \nabla^{2}) \mathbf{U} \nonumber \\ 
& + \left[ -\frac{1}{\mu} \bsf{I} + \frac{\tau}{\mu} (\alpha + 1 ) \bsf{S} + \tau (\nabla \mathbf{F})^{\intercal}\right] \cdot \mathbf{v} \nonumber \\ 
& + \frac{\tau \beta}{\mu} \sqrt{2D} \mathbf{\eta}, 
\label{eqn:generalizedF}
\end{align}
where the first two terms on the right hand side present a generalized reversible force, the third term is a generalized velocity-dependent force and the last term contains the noise. Thus \eqref{eqn:generalizedF} becomes: 
\begin{widetext}
\begin{equation}
\label{eqn:inertialEoM}
\frac{\tau}{\mu} \dot{\mathbf{v}}  = \left[-\frac{1}{\mu}\bsf{I} + \frac{\tau}{\mu}(\alpha + 1)\bsf{S}  \right]\cdot({\bf v} - {\bf U}) + {\bf F_{eff}} + \frac{\tau}{\mu}(D_t - \beta \ell^2 \nabla^2){\bf U} + \frac{\tau \beta}{\mu} \sqrt{2D} \bm{\eta} 
\end{equation}
\end{widetext}
where $D_t = \partial_t + {\bf U} \cdot \nabla$, $\bf F_{eff} = \bf{F} + \tau \dot{\bf{F}} - \tau \left( \alpha\bsf{S} + \bsf{A} \right)\cdot\bf{F} $, is the effective external force on the particle, and all fields are evaluated at $\vect{X}(t)$. In the absence of $\ell^2 \nabla^2\mathbf{U}$ and $\bm{\eta}$ \eqref{eqn:dimerw} is homogeneous, so that $\beta$ can be absorbed into $\mathbf{w}$ in \eqref{eqn:dimer}. This is why $\beta$ appears only as a prefactor of $\nabla^2 \mathbf{U}$ and $\bm{\eta}$ in \eqref{eqn:inertialEoM}. However, \eqref{eqn:dimer} can be recast as \eqref{eqn:inertialEoM} only if $\mathbf{w}$, i.e., self-propulsion, enters \eqref{eqn:dimerX}; the degree to which it does so is governed by $\tau$ in \eqref{eqn:dimerw}. For $\tau \to 0$, \eqref{eqn:dimerw} implies $\mathbf{w} = 0$. 

The presence in \eqref{eqn:inertialEoM} of the external force and the drag, unmodified by prefactors, means that $\tau/\mu$ plays the role of an effective mass for this inertia-less active system. Indeed \eqref{eqn:inertialEoM} resembles the Maxey-Riley equation for inertial particles in a flow \cite{MR1983}. There are differences, such as the absence in \eqref{eqn:inertialEoM} of the Basset-Boussinesq history term \cite{MR1983,prasath2019accurate}, but the %intriguing 
similarities prompt us to explore analogs to passive inertial-particle behavior in the dynamics of active inertia-less particles in external flows.

\section{Caustics near a point vortex flow: singular perturbation analysis}\label{sbp}
We begin with the classical case of motion in the flow field of a point vortex. In plane polar coordinates $(r,\theta)$, $\vect{U} = \hat{\theta} \tilde{\Gamma}/r $, with circulation $2 \pi \tilde{\Gamma}$. Note that $\bsf{A} = 0$ for this flow everywhere except at the origin. To demonstrate caustics even for simple fore-aft symmetric particles in the absence of external forcing and noise, we set $\ell = 0$, $\eta = 0$, and  $\bf{F} = 0$. Non-dimensionalizing \eqref{eqn:inertialEoM} using the natural length $\sqrt{\tilde{\Gamma} \tau}$ \cite{Croor2015} and time $\tau$, we get the coupled equations
\begin{subequations}
\label{eqn:rLeqns}
\begin{equation}
\ddot{r} - \frac{L^{2}}{r^{3}} = - \dot{r} + \frac{\alpha}{r^{3}} - \frac{(1+\alpha) L}{r^{3}},
\label{eqn:reqn}
\end{equation}
%\begin{equation}
%\dot{L} = 1 - L - \frac{(1+\alpha) \dot{r}}{r} + %\frac{\tilde{l}}{r^{2}},
%\label{eqn:5.9}
%\end{equation}
\begin{equation}
\dot{L} = 1 - L - \frac{(1+\alpha) \dot{r}}{r}, %+ \frac{\lambda}{r^{2}}
\label{eqn:Leqn}
\end{equation} 
\end{subequations}
where the overdot denotes $d/dt$. Eqs. \eqref{eqn:rLeqns} govern the Lagrangian dynamics of an active particle whose centroid is at a radial distance of $r$ from a point vortex. Here, $L \equiv r^{2} \dot \theta$ is the angular momentum per unit mass of the particle. For a detailed derivation of \eqref{eqn:rLeqns}, visit Appendix \ref{AOUPvortex}.

An effective centrifugal acceleration, reinforcing the similarity to an IP, arises through the $L^2/r^3$ term. Due to the terms containing $\alpha$, our equations are distinct from those for true IP \cite{Croor2015}. We treat the nonlinearities in \eqref{eqn:reqn} \& \eqref{eqn:Leqn} perturbatively. A \textit{regular} perturbation approach yields absurd solutions near the origin; in fact equations \eqref{eqn:reqn} and \eqref{eqn:Leqn} constitute a singular perturbation problem \cite{BO1999}.
The behaviour at very small times and small distances away from the vortex is singular, and relatively violent, unlike the more gentle relaxation to the final state at late times. We exploit this feature to understand the different physics at small and large time. %\remarkSR{Need to talk to you to understand this better.}

%\begin{figure}[t]
%      \begin{center}
%      \includegraphics[width=8.5 cm]{fig2.jpg}
      %\hspace{0cm}\newline
%      \caption{\label{centrifugation}\textbf{Centrifugation and caustics of transient active dimers (noiseless AOUP)}: (a) Shows time frames showing the positions of the particles (blue dots) around a point vortex at the origin for IP and noiseless AOUP respectively [see Supplementary video 3.1 \& 3.2]. Particles were initialised with uniformly random initial positions and orientations/velocities. (b) An initially homogeneous number-density (grey circles) peaks near a critical radial distance (green) in the steady state ($t \gg 1$), compared to unsteady density of IP (purple) at the same time. (c) Trajectory (rays) of particles averaged over all initial orientations. The envelope of rays, for particles starting at various initial $r$,  gives rise to the caustics (red circles); and $R = t^{\nu}$ (green curve), with $\nu =2/3$. In (b), (c) \& (d), $\alpha=1$.} 
%      \end{center}
%\end{figure}

We seek an inner solution at the lowest order for $t\ll 1$ and $r\ll 1$, and an outer solution for $t\gg1$, where $r$ could be $O(1)$ or larger. In contrast to an IP which centrifuges out forever, the outer solution for an %AOUP 
active Hookean dimer is steady rotation with $\mathbf{w}$ and $\dot\theta=1/r_f^2$ at a constant final distance $r_f$ from the vortex, rendering the particle passive  %\remarkSR{[it's anyway without inertia so I took out ``lifeless and inertialess''.]} %and inertialess 
at large times. The unexpected %Intriguing 
physics that appears in the inner region %, which 
sets the stage for the rest of this article. 

\subsection{Inner solution, $r \ll 1$} 
A particle starting near the vortex origin spends little time in its vicinity, centrifuging out quickly to large distances. We may write a dominant balance equation applicable to this region. As is standard in singular perturbation theory, we recast \eqref{eqn:reqn} and \eqref{eqn:Leqn} in the stretched spatial $\tilde{r} \equiv r/\delta_i$ and temporal $\tilde{t} \equiv t/\epsilon_i$ variables, where $\epsilon_i$ and $\delta_i$ are as yet unknown, but will be selected to ensure that all derivatives in the stretched variables are $\mathcal{O}(1)$. This rescaling preserves the angular momentum per unit mass, $\tilde{L}  = L$. Examining equation \eqref{eqn:reqn} and \eqref{eqn:Leqn} tells us that, for centrifugation to occur, the time derivatives must be much larger than $O(1)$ in the $r \to 0$ limit. Dominant balance gives the following asymptotic equation for $L$, 
\begin{equation}
\dot{L} = - \frac{(1+\alpha) \dot{\tilde{r}}}{\tilde{r}},
\label{eqnL}
\end{equation} 
with solution $L(\tilde{r}) = L_0 -(1+ \alpha)\log{\tilde{r}/\tilde{r}_0}$, where $L = L_0$ when $r = r_0$, a constant. The subscript $0$ indicates an initial value. Similarly, the equation for $\tilde{r}$ becomes 
\begin{align}
\label{smallrL}
\begin{split}
        \frac{\delta_i}{\epsilon_i^2}\left( \ddot{\tilde{r}}  - \frac{L^2}{\tilde{r}^3} \right) &= \frac{\delta_i}{\epsilon_i}\dot{\tilde{r}} + \frac{1}{\delta_i^3\tilde{r}^3} \left( \alpha - \frac{\delta_i^2}{\epsilon_i}(\alpha + 1)L\right). 
\end{split}
\end{align}
%The above 
which immediately provides the relationship $\epsilon_i \sim \delta_i^{2}$ between the two small quantities. 
Making the choice $\epsilon_i = \delta_i^{2}$ in \eqref{smallrL} leads, at $\mathcal{O}(1)$, to the autonomous equation 
\begin{equation}
\tilde{r}^3 \ddot{\tilde{r}} = (L-1)(L-\alpha).
\label{smallrr}
\end{equation} 
%which is an autonomous equation. 
Equations \eqref{eqnL} and \eqref{smallrr} yield the dynamics $\dot{\tilde{r}} = \partial \mathcal{H}/\partial p$ and $\dot{p} = -\partial \mathcal{H}/\partial \tilde{r}$ with an effective Hamiltonian
\begin{equation}
\mathcal{H} = \frac{p^{2}}{2} + \frac{\alpha + [L(\tilde{r}) - 1 -\alpha]^2}{2 \tilde{r}^2}
%v_{r}^{2} = {v_{0}}^{2} -  \frac{\alpha + (L - 1 -\alpha)^2}{\tilde{r}^2} +
%\frac{\alpha + (L_0 - 1 -\alpha)^2}{\tilde{r}_0^2},
%\frac{f(\tilde{r},\tilde{r}_0,r^*)}{\tilde{r}^2} + 
% \frac{f(\tilde{r},r^*)}{\tilde{r}^2} +
% \frac{f(\tilde{r}_0,r^*)}{\tilde{r}_0^2},
\label{eqn:5.13}
\end{equation}
where $p \equiv d\tilde{r}/d\tilde{t}$ is the radial momentum per unit mass of the particle. Hamiltonian structure in an overdamped dynamical system has been observed earlier in self-propelled \cite{Stark2016, Lushi2015, Shelley2019, Arguedas_Leiva_2020, Ronojoy2020} and externally driven \cite{Chajwa2019, Chajwa2020} particles. Integrating equation \eqref{smallrr}, using the chain rule $\ddot{\tilde{r}} = p dp  / d\tilde{r}$ and the inner solution $L(\tilde{r})$ of \eqref{eqnL} gives the constant-energy manifolds  
\begin{equation}
p^{2} = {p_{0}}^{2} -
\frac{\alpha + (L - 1 -\alpha)^2}{\tilde{r}^2} +
\frac{\alpha + (L_0 - 1 -\alpha)^2}{\tilde{r}_0^2},
\label{eqn:5.13b}
\end{equation}
in the effective $(\tilde{r}, p)$ phase space, where each level set corresponds to a ray in the $(t,r)$ plane, acquired by integrating \eqref{eqn:5.13b}. 
We start with two rings of particles around the vortex at initial radii $r_0$ and $r_0+\Delta r$. The intersection of their rays in the $r,t$ plane represents an overtaking of the outer ring by the inner, i.e., the occurrence of caustics.

Caustics for a few chosen $r_0$ are shown by the blue circles in Fig.\ref{Figure1}(c), along with the envelope (continuous line) for all $r_0$, representing the smallest radial distance at a given time for the occurrence of caustics. For the purpose of demonstration we have taken $L_{0}= - (1+\alpha) \log \tilde{r}_{0}$, and equal initial speeds. This picture is akin to geometrical optics, where caustics form an envelope tangent to the light rays \cite{eggers_fontelos_2015}. 

\begin{figure}[t]
\centering
\includegraphics[width = 0.5 \textwidth]{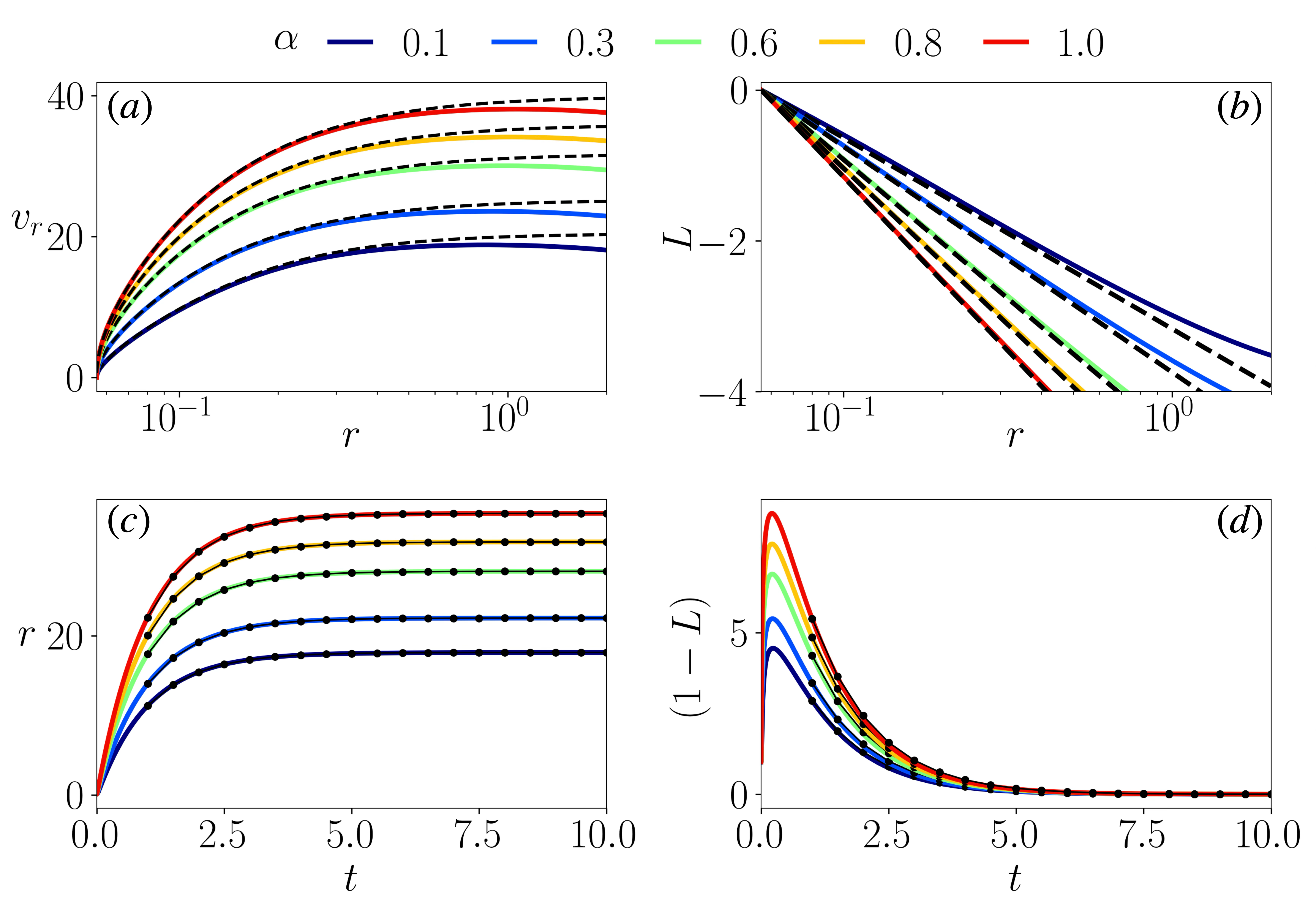}

\caption{\textbf{Comparison between asymptotic analysis and full numerical solution:}(a,b) Comparing trajectories obtained analytically from the inner solution in \eqref{eqn:5.13b} (dashed lines), and from evolving the full system following \eqref{eqn:rLeqns} (curves). Both the radial velocity in (a) and angular momentum in (b) follow the analytical solution till particles reach $O(1)$ distance from the vortex origin, where the inner solution is no longer valid. (c,d) The outer solution in black dotted lines following \eqref{eqn:osol} also matches well with the numerical solution of the full system at late times. At such times, particles are far away from the vortex origin (c), and their angular momentum saturates to the background fluid angular velocity (d).}\label{io}
\end{figure}

Panels (a) and (b) in Fig. \ref{io} compare the inner solution with the full solution obtained by numerically time-integrating \eqref{eqn:rLeqns} in the form of particle trajectories for different $\alpha$s averaged over initial particle velocity. 
%\remarkSR{[I'm suddenly confused, maybe not enough sleep: why is one averaging over initial velocities?]} 
For all activity values, the inner solution in dashed lines matches well when $r \ll 1$. Around $r$ of order unity, the inner solution expectedly deviates from the numerical solution as higher order corrections of $\epsilon_i$ become significant. Following Fig. \ref{io}(a), we see that particles with higher activity centrifuge out with larger velocities. This is expected from \eqref{eqn:reqn}, as $\alpha$ is involved in terms that have $r^3$ in the denominator, which control centrifugation. $\alpha$ also controls the slope in \eqref{eqn:Leqn}, and therefore we see $L$ decreases faster with larger $\alpha$ in Fig. \ref{io}(b). 

% \begin{figure}[!h]
%       \begin{center}
%       \includegraphics[width=8 cm]{inner_caus.jpg}
%       %\hspace{0cm}\newline
%       \caption{\label{Fig52} Caustics of the inner solution is marked by the intersection of trajectories (purple circles) starting at a closely separated initial radial distance $\Delta r_0 = 0.05$, and $\alpha = 1.0$. The limit $\Delta r_0 \to 0$ would give a curve of purple circles. }
%       \end{center}
% \end{figure}
\begin{figure}[t]
      \begin{center}
      \includegraphics[width=8.5 cm]{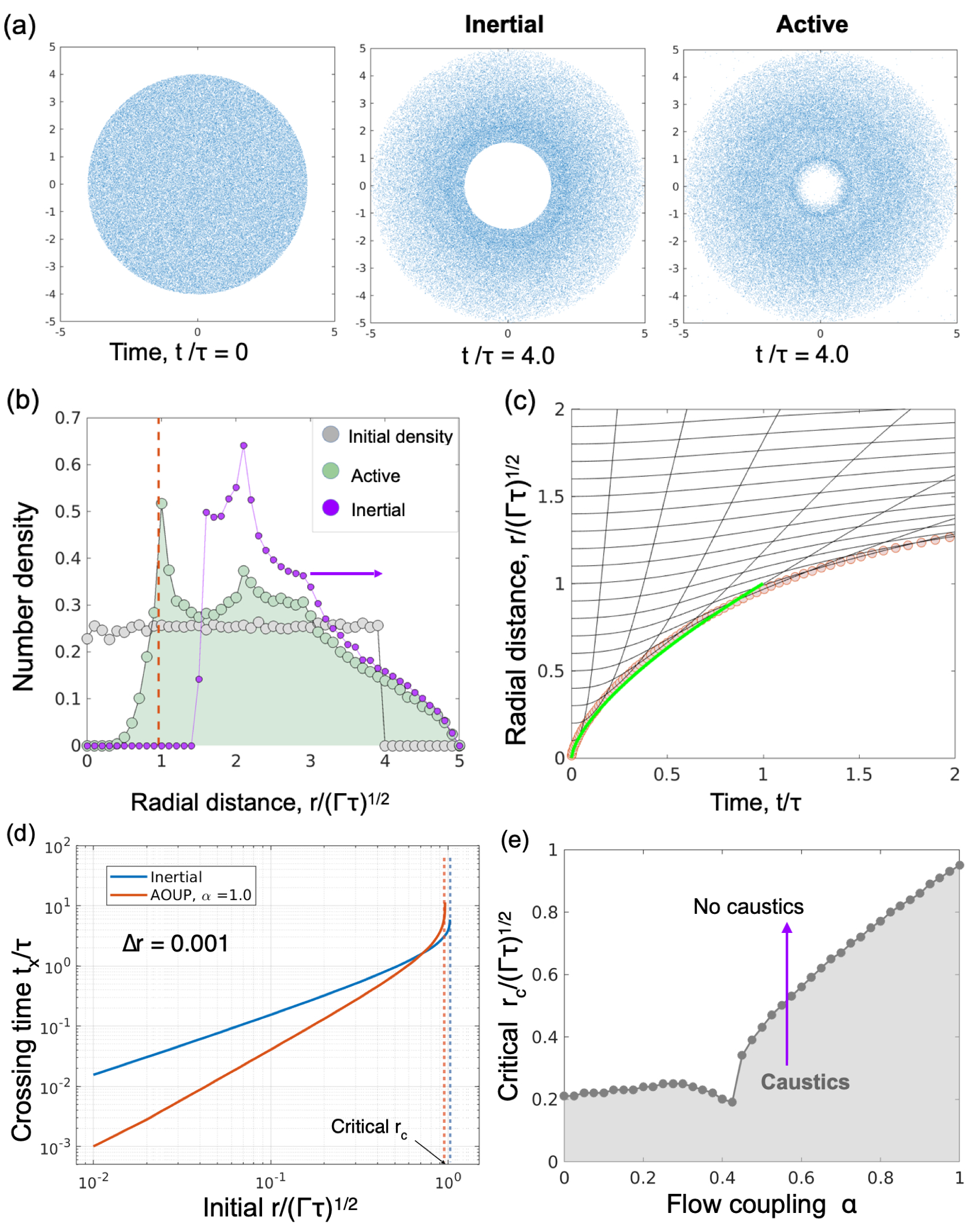}
      %\hspace{0cm}\newline
      \caption{\label{caustics_regime} \textbf{Centrifugation and caustics of a Hookean dimers in point vortex}:  (a) Time frames showing the positions of the particles (blue dots) around a point vortex at the origin for IP and noiseless AOUP, by numerically solving the Maxey-Riley equation, and \eqref{eqn:dimerX} \& \eqref{eqn:dimerw}, respectively [see Appendix \ref{videos} Video 1]. $N=10^{5}$ particles were initialised with uniformly random initial positions and orientations/velocities. (b) An initially homogeneous number-density (grey circles) peaks near a critical radial distance (green) in the steady state, compared to unsteady density of IP (purple) at a representative $t \gg 1$. (c) Trajectory (rays) of particles averaged over all initial orientations. The envelope of rays, for particles starting at various initial $r$, gives rise to caustics (red circles). The green curve shows $R = t^{\nu}$, with $\nu =2/3$. In (b), (c) \& (d), $\alpha=1$. (d) 
      The crossing time of adjacent rays separated by $\Delta r = 0.001$ starting at various radial distances $r$, averaged over uniformly random initial orientations, diverges at a finite critical radius $r_c$ for an active particle (red) and for an inertial particle (blue) averaged over uniformly random initial velocity of unit magnitude. (e) For various $\alpha$, plotting $r_c$ demarcates the region of caustics, which is the radial distance below which adjacent rays cross each other in a finite time. } 
      %(d) Caustics phase diagram in the $R_0-\alpha$ plane.}
      \end{center}
\end{figure}

\subsection{Outer solution, $r\gg 1$}
At long times, particles move far away from the vortex origin i.e. $r\gg 1$ for $t \gg 1$. This motivates us to rescale $r$ and $t$ in \eqref{eqn:rLeqns} such that the rescaled outer variables $\tilde{t} = t\epsilon_o $ and  $\tilde{r} = r\delta_o$, are $\mathcal{O}(1)$ for small $\epsilon_o$ and $\delta_o$. Substituting this in \eqref{eqn:reqn} gives a constant $\tilde{r}(t)$, with no caustics, and no activity. Thus, there is a critical initial radial distance $r_0$ above which particles asymptotically approaches a final radius $r_f$, without caustics with final angular momentum per unit mass, $L_f  = 1$. To explore this dynamics in detail, we expand $r$ in series of $\delta_o$, 
\begin{align}
% \label{}
\begin{split}
    r(t) = \frac{\t{r}_f}{\delta_o} - \t{r}(t) + \mbox{O}(\delta_o) ,
\end{split}
\end{align}
where $\t{r}_f = r_f\delta_o$, and $\delta_o \ll 1$. The non-dimensional equation \eqref{eqn:rLeqns} can be recast as
\begin{align}
% \label{}
\begin{split}
    -\ddot{\t{r}}  &= \dot{\t{r}} + \mbox{O}(\delta_o^3),\\ 
    \dot{L} &=  1 - L + \frac{\delta_o(\alpha  +1) }{\t{r_f}}\dot{\t{r}} + \mbox{O}(\delta_o^2)\ . 
\end{split}
\end{align}
Using the boundary conditions $r(t^*) = r^*, \dot{r}(t^*) = v_r^*$, and $L(t^*) = L^*$, we obtain the leading order solution of $r$ and $L$,
\begin{subequations}

\begin{equation}
\begin{split}
L = 1 - (1-L^*(\alpha))e^{-(t -t^*)} +\left( t - t^* \right) \\+\left( \alpha + 1 \right)\frac{v_r^*}{r_f}\,  e^{-(t -t^*)}
 + \mbox{O}(\delta_o^2), 
\end{split}
 \end{equation}
 \begin{equation}
r = r_f - v_{r}^{*}e^{-(t -t^*)} + \mbox{O}(\delta_o^2)  \ .
\end{equation}
\label{eqn:osol}
\end{subequations}
For every particle, $r_f, L^*$, and $v_r^*$ vary with $\alpha$, as shown in Fig. \ref{io} (c) and (d) which compares the above outer solution to the full numerical solution for various values of $\alpha$. Choosing the $t^*$ carefully such that the particle is far away from the origin, we see that the outer solution in dotted lines matches well with the full numerical solution. The inner and outer solutions uncover two qualitatively distinct dynamics, one near and the other far away from the vortex origin. The inner solution presents rapid centrifugation and caustics, whereas the outer solution yields exponential relaxation of the particle trajectories to the background flow. We explore the caustics in the inner region and their dependency on $\alpha$ in the next section.

\section{Active Dimer Cloud in a Point Vortex: Numerical Solution}
\label{hookeancloud}
We now numerically solve the full equations \eqref{eqn:dimerX} and \eqref{eqn:dimerw}, with the noise, the external force ${\bf F}$, %potential, 
and the polar flow coupling $\lambda$ set to zero (see Appendix \ref{videos} Video 1). In this case the self-propulsion $\beta$ can be absorbed in the definition of $\vect{w}$. 
Early in the process, for times $t$ less than or comparable to $\tau$, we find that noiseless AOUP, i.e., active Hookean dimers, behave similar to IP, in that they both display centrifugation close to the vortex [see middle panel of Fig. \ref{caustics_regime}(a)]. Such voiding of vortical regions has been observed previously 
for rigid gyrotactic swimmers \cite{Durham2011} and bacteria \cite{Sokolov2016}, and is shown to be critical for their transport in turbulent environments \cite{Zhan2014,Torney2007}. 
However, the identification of a singularity, namely caustics, for this flow- and motility-induced dynamics was missing. At long times, the radial profile of number density reaches a steady state, with a peak at a particular radius [Fig. \ref{caustics_regime}(b)] 
The extension of the dimer relaxes to zero, so in the absence of noise, the AOUP at long time behaves like a tracer particle, consistent with our asymptotic analysis. In contrast, IPs centrifuge out forever, though more slowly as time progresses [see Appendix \ref{videos} Video 1] since the drag force balances the centrifugal force and radial velocity decays as $r^{-3}$ \cite{Croor2022}. This feature of IPs is closely mimicked by a more robust 
motile particle [see Appendix \ref{videos} Video 2], which we discuss below.
The sharp peak in the number density at a particular radial distance coincides with the formation of caustics, which we obtain by averaging 
over uniformly random initial orientations. The caustics are seen in Fig. \ref{caustics_regime}(c) with an envelope as predicted by the inner solution in the paragraph leading up to \eqref{eqn:5.13}. The scaling of this caustics curve near the vortex singularity is $r \sim t^{2/3}$ [green curve in Fig. \ref{caustics_regime}(c)], similar to the power-law scaling of the optical caustics near its singular tip \cite{eggers_fontelos_2015} [Fig. 1(c) inset], and distinct from the scaling in IP. 

We demarcate the regions in the $r_0-\alpha$ plane where caustics occur as those where an intersection of adjacent rays (see Fig. \ref{caustics_regime}(e)) takes place in finite time, $t_x$. 
At a particular $r_0=r_c$, $t_x$ diverges [see Fig. \ref{caustics_regime} (d)], and caustics do not occur when the initial particle position is beyond this critical radius. %This is seen to be comparable to that for IP \cite{Croor2015}. 
This behaviour is similar to that of IPs \cite{Croor2015}. 
The flow coupling $\alpha$ has a dual role at small to moderate distances from the vortex: (i) it aligns the dimer along the stable principal axis of $\bsf{S}$, which has a non-zero radial component; (ii) once aligned, it extends the dimer along this axis, thus competing with the $1/\tau$ relaxation to zero motility. However, when the dimers have been centrifuged out to large $r$, the relaxation term takes over, leading to tracer-like dynamics, and the caustics radii lie at intermediate values of $r$ [see Fig. \ref{caustics_regime} (b)].

\section{Caustics from the continuum limit of active particles} \label{Continuum}
Thus far, we have obtained caustics by evolving pairs of particles close to each other and then checking if their paths cross, where the initial separation between the two particles, $\Delta r$, is a free parameter that is fixed arbitrarily. In this section, by turning to an alternative formulation of caustics, we establish that our choice of $\Delta r$ does not affect our answers. We transition from a discrete particle system to a continuous particle phase, an approach common for studying IP caustics \cite{falkovich2002,meibohm2021,PRL2023}. Consider a continuum of Hookean active dimers, each obeying \eqref{eqn:inertialEoM} without noise. Their dynamics can be described by a velocity field $\vect{v}=\vect{v}(\vect{x},t)$, whose equation follows directly from \eqref{eqn:inertialEoM}, 
%\remarkSR{[we haven't explained where this equation comes from]}
% \begin{align}
%         \label{eom_act}
%         \frac{\tau}{\mu}d_t v_i & = -\left[ \frac{1}{\mu} - \frac{\tau}{\mu}\left( \alpha + 1 \right)%\bsf
%         \sf{S}_{ik} \right] \left( v_k - U_k \right) + (\textnormal{F}_{\textnormal{eff}})_i  \nonumber \\
%         & + \frac{\tau}{\mu}\left( D_t - \beta \ell^2 \nabla^2  \right)U_i + \frac{\beta}{\mu}\sqrt{2D}\,\eta_i \ ,
%     \end{align}
\begin{align}
        \label{eom_act}
        \frac{\tau}{\mu}d_t v_i & = -\left[ \frac{1}{\mu} - \frac{\tau}{\mu}\left( \alpha + 1 \right)%\bsf
        \sf{S}_{ik} \right] \left( v_k - U_k \right) + (\textnormal{F}_{\textnormal{eff}})_i  \nonumber \\
        & + \frac{\tau}{\mu}\left( D_t - \beta \ell^2 \nabla^2  \right)U_i ,
    \end{align}
using the Einstein summation convention, where $d_t \equiv \partial_t  + v_k\partial_k$ is the particle-phase material derivative. 
%where repeated indices are summed over. 
Note that \eqref{eom_act} is identical in form to \eqref{eqn:inertialEoM} except in the definition of $d_t$. %\rishi{Similar transition from discrete particle system to continuous particle phase is common for studying IP caustics \cite{falkovich2002,meibohm2021,PRL2023}.} 
%\remarkRC{$\eta$ in \eqref{eqn:inertialEoM} is a function of time, while in (12) its a function of both space and time. Is the spatial gradient of $\eta$ well defined? Differentiability of $\eta$ in (12) in space is not obvious from (3).}
The particle velocity field generally has a non-zero divergence since the particle flow is compressible. However, it is continuous everywhere except in caustics regions where this divergence $\vect{\nabla}\cdot\vect{v}\to -\infty$ \cite{meibohm2d2021}, that is, $\Tr(\bsf{Z}) \to -\infty$ where $\bsf{Z} =\tau\grad{\vect{v}}$ is a non-dimensional particle-velocity gradient tensor. %An equation for $\bsf{Z}$ readily follows from equation \eqref{eom_act}
The dynamics of $\bsf{Z}$ readily follows from equation \eqref{eom_act},
    % \begin{widetext}
    % \begin{align}
    %     \label{zevol_act}
    %         d_tZ_{ij} = &-\frac{1}{\tau}\left( Z_{ij} - \tilde{O}_{ij} +Z_{ik}Z_{kj} \right)+ \mu\partial_j(\textnormal{F}_{\textnormal{eff}})_i + \left( D_t - \beta \ell^2 \nabla^2  \right)\tilde{O}_{ij} +\frac{1}{\tau}\tilde{O}_{ik}\tilde{O}_{kj} + \beta\sqrt{2D}\,\partial_j\eta_i \nonumber \\
    %         &+ \frac{ \alpha + 1}{2}\left[ \left( v_k -U_k \right).\partial_k
    %         {\tilde{O}_{ij}} + \left( v_k - U_k \right)\partial_i\tilde{O}_{kj} + 2S_{ik}\left( Z_{kj} - \tilde{O}_{kj}\right) \right] ,
    %     \end{align}
    % \end{widetext}
    \begin{widetext}
    \begin{align}
        \label{zevol_act}
            d_tZ_{ij} = &-\frac{1}{\tau}\left( Z_{ij} - \tilde{O}_{ij} +Z_{ik}Z_{kj} \right)+ \mu\partial_j(\textnormal{F}_{\textnormal{eff}})_i + \left( D_t - \beta \ell^2 \nabla^2  \right)\tilde{O}_{ij} +\frac{1}{\tau}\tilde{O}_{ik}\tilde{O}_{kj} \nonumber \\
            &+ \frac{ \alpha + 1}{2}\left[ \left( v_k -U_k \right).\partial_k
            {\tilde{O}_{ij}} + \left( v_k - U_k \right)\partial_i\tilde{O}_{kj} + 2S_{ik}\left( Z_{kj} - \tilde{O}_{kj}\right) \right] ,
        \end{align}
    \end{widetext}
where $\tilde{\bsf{O}} =\tau\grad{\vect{u}} =  \tau\left(  \bsf{S} + \bsf{A}\right)$ is the non-dimensional fluid gradient tensor with $\tilde{\bsf{S}} = \tau \bsf{S} $ and $ \tilde{\bsf{A}} = \tau\bsf{A}$ being its symmetric and anti-symmetric part, respectively. The benefit of equation \eqref{zevol_act} is that it can be solved in the Lagrangian frame of individual particles and can thus predict whether a given particle will experience a caustic. %\remarkRC{is it "a particle" or a parcel of particles undergoing caustics?} 
Furthermore, $\bsf{Z}$ at a single point quantifies the differences in particle velocities in the neighbourhood of that point.

    \begin{figure}[b]
        \centering
        \includegraphics[width = \linewidth]{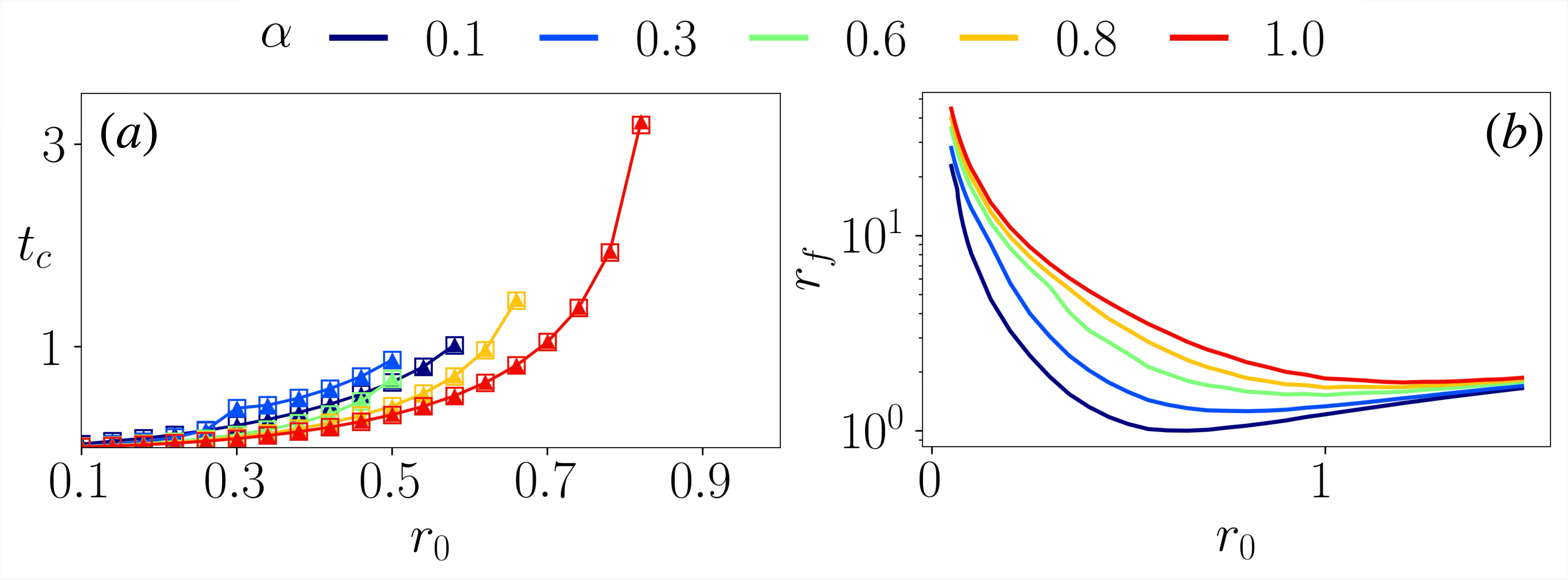}
        \caption{  (a) Caustics times for particles starting from different initial radii as calculated from equations \eqref{nd_prtclEvol} (triangles) and \eqref{nd_ZEvol} (squares). The two definitions of caustics give the same answer. (b) Average final radius for a range of initial radius, for different activity levels $\alpha$, of particles starting near a point vortex. }\label{pt_vortex}
    \end{figure}
In the following, we show caustics formation by active Hookean Dimers near a point vortex using both equations \eqref{eom_act} and \eqref{zevol_act} and confirm that they both yield the same result. We consider fore-aft symmetric particles without external forcing i.e., $\ell = \bsf{F}_{\textnormal{eff}} = 0$, just as in Section \ref{sbp}. Using the characteristic length and timescale of $\sqrt{\tilde{\Gamma \tau}}$ and $\tau$, respectively, equation \eqref{eom_act} reduces to the non-dimensional form
    \begin{align}
    \label{nd_prtclEvol}  
    \begin{split}
        d_tv_i = -\left[ 1 - \left( \alpha + 1 \right)\tilde{S}_{ik} \right] \left( v_k - U_k \right) + D_t U_i  \ ,
    \end{split}
    \end{align}
where, for a point vortex 
%in the cartesian coordinate system $(x,y)$ takes the form
    \begin{align}
        % \label{}
        \begin{split}
            \tilde{\bsf{S}} \equiv -\frac{1}{r^2}\left[
                \begin{array}{cc}
                    -\sin 2\theta & \cos2\theta \\
                    \cos2\theta & \sin 2\theta
                \end{array}
                \right] = \tilde{\bsf{O}}\ ,
            \end{split}
    \end{align}
which is well defined everywhere except at the origin, with $\theta = \tan^{-1}y/x$ in terms of Cartesian $x,y$ coordinates.
Similarly, equation \eqref{zevol_act} becomes
    \begin{align}
    \label{nd_ZEvol}
    d_t\bsf{Z} = & -\left( \bsf{Z} - \tilde{\bsf{S}} +\bsf{Z}^2 \right)+   D_t \tilde{\bsf{S}} +\tilde{\bsf{S}}^2 \nonumber \\ 
    & + \left( \alpha + 1 \right)\left[ \left( \vect{v} -\vect{U} \right).\grad{\tilde{\bsf{S}}} + \tilde{\bsf{S}}\cdot\left( \bsf{Z} - \tilde{\bsf{S}}\right) \right] \ .
    \end{align}
    For a stationary background flow, such as a point vortex,
 \begin{align}
        % \label{}
        \begin{split}
        D_tU_i = \vect{U}.\grad{U_i}\quad {\rm and} \quad D_t\tilde{\bsf{S}} = \vect{U}.\grad{ \tilde{\bsf{S}}}.
    \end{split}
\end{align}
With these equations, we initialize active Hookean particles at different radii, with initial velocity $\vect{v}(0) = \sin(r) \vect{U}$, and evaluate where and when they form caustics. Fig. \ref{pt_vortex} (left) shows the time taken for different particles to form caustics starting from a radius $r_0$ for different values of $\alpha$. The caustics detected by the $\Tr{\bsf{Z}} \to -\infty$ (squares) practically overlap with those obtained from crossing of particle trajectories (triangles).

Another way of looking at whether particles starting from a particular radius formed caustics is the following. If two rings of particles start with initial radii $r_{01}$ and $r_{02}$ with $r_{01}< r_{02}$ and end at final radii $r_{f1}$ and $r_{f2}$ respectively, with $r_{f1}>r_{f2}$, they must have undergone caustics during their evolution. We can thus identify, from the regions of negative slope, that caustics occur almost everywhere in the range of initial locations shown in Fig. \ref{pt_vortex} (right) for particles with different activity with $r_0 \ll 1$.
% \begin{figure}[H]

\section{Active Preferred-Length Dimer in a Point Vortex}
    
Although Active Hookean dimers are analytically tractable and offer a conceptual understanding of the coupling between flow and the activity of deformable swimmers, their extension, and hence their intrinsic speed, relax to zero in the absence of noise and flow. The motility parameter $\beta$ defines a speed only when multiplied by a preferred scale of ${\bf w}$, say its RMS value. In nature, motile organisms generally possess an intrinsic speed independent of noise. To study such a case, we consider the equations for position $\mathbf{X}$ and extension $\mathbf{w}$ for an active dimer with a strongly preferred value of $|\mathbf{w}| = w_{0}$ and speed $v = \beta w_{0}$,
\begin{equation}
\dot{\mathbf{X}}  =  \mu \mathbf{F} + \mathbf{U} + \beta 
{\mathbf{w}}% \equiv \vect{v},   
\label{eqn:dimerX5}
\end{equation}
\begin{align}
\dot{\mathbf{w}} & = \frac{1}{\tau} \left(1 - \frac{|\mathbf{w}|^2} { w_0^2}\right) \mathbf{w} \, + \, (\alpha \bsf{S} + \bsf{A})\cdot \mathbf{w} \nonumber \\ &- {\ell}^{2} \nabla^{2} \vect{U} \, %+ {\sqrt{2D} \over \tau}\bm{ \eta}(t)\label{eqn:dimerX6}.
+ \sqrt{2D}\bm{ \eta}(t).\label{eqn:dimerX6}
%{\sqrt{\frac{2D}{U_{0}^2 \tau}}}
\end{align}
Equations \eqref{eqn:dimerX5} and \eqref{eqn:dimerX6} describe an active Brownian particle (ABP) in a flow. Note that instead of  a projection operator $\mathbb{I} - \vect{w}\vect{w}$ on $(\alpha \bsf{S} + \bsf{A})\cdot \mathbf{w}$ that imposes a rigidity constraint \cite{Jeffery1922}, the $\tau$ dependent term %\rgc{Not clear what  $\tau$ dependent relaxational term is, since tau is relaxation time } 
in \eqref{eqn:dimerX6} allows soft distortions about the preferred length $w_{0}$. This difference between our preferred-length model and the traditional ABP, in which $|{\bf w}|$ is constant, is unimportant. 
\begin{figure}[t]
      \begin{center}
      \includegraphics[width=8.5 cm]{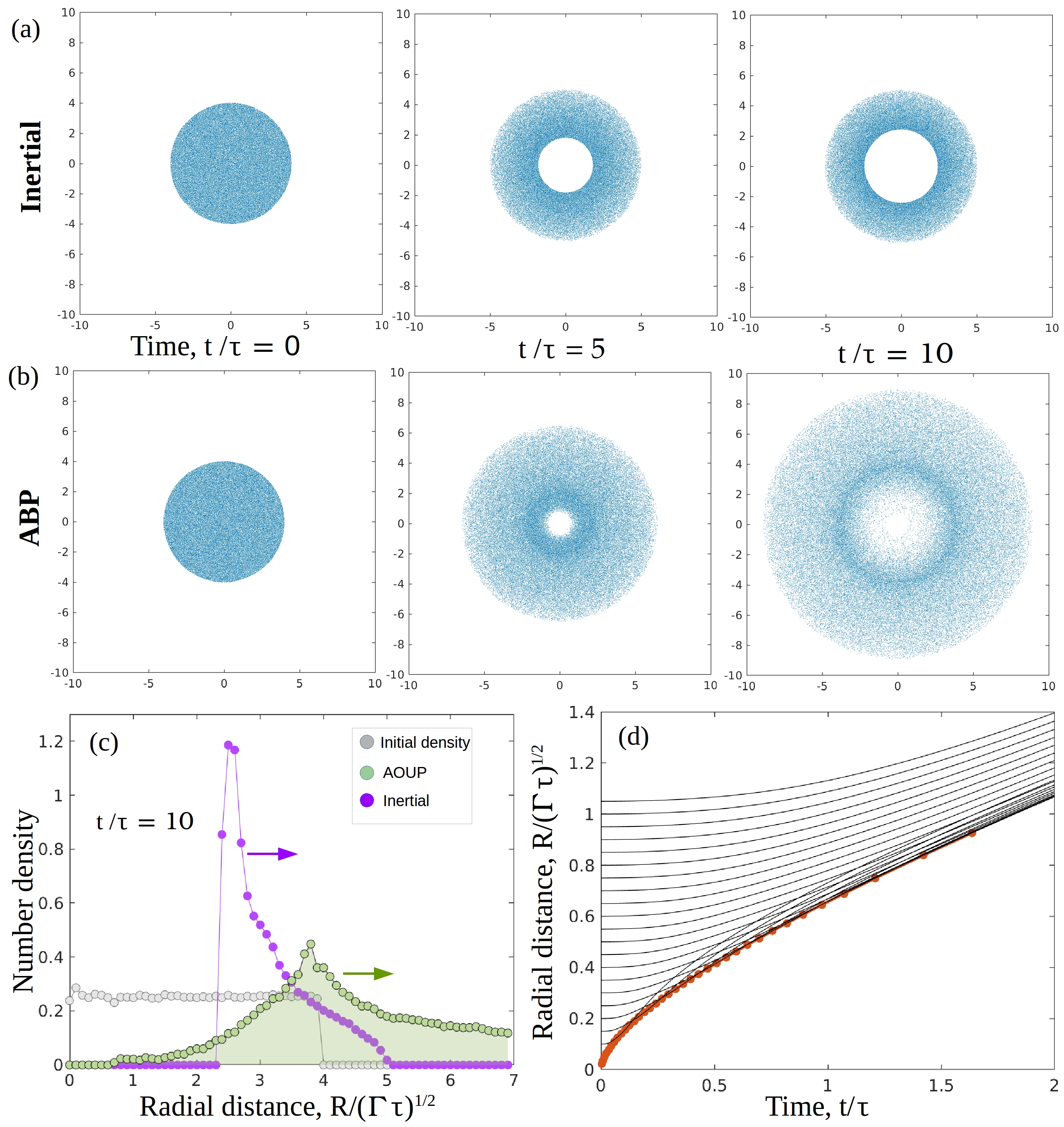}
      %\hspace{0cm}\newline
      \caption{\label{fig:abp1}\textbf{Centrifugation and caustics of active preferred-length dimers in a point vortex}: (a) \& (b) show time frames of particle positions around a point vortex for inertial particles and preferred-length dimers (or ABP without noise) respectively, shown for $\tilde{\beta} = 0.5$ and $\alpha=1.0$ in \eqref{eqn:dimerX5} - \eqref{eqn:dimerX6}. (c) radial number-density of active particles (green) compared with inertial particles (purple) shown for time $10\tau$, and the number-density at $t=0$ (grey). Arrows schematically depict the radial drift of the unsteady state. (d) Intersections of adjacent trajectories marking the caustics curve (red circles) obtained for $\tilde{\beta} = 0.5$, $\alpha = 1.0$ and $\Delta r = 0.01$.} 
      %(d) Caustics phase diagram in the $R_0-\alpha$ plane.}
      \end{center}
\end{figure}

\begin{figure*}[t]
          \begin{center}
          \includegraphics[width=18 cm]{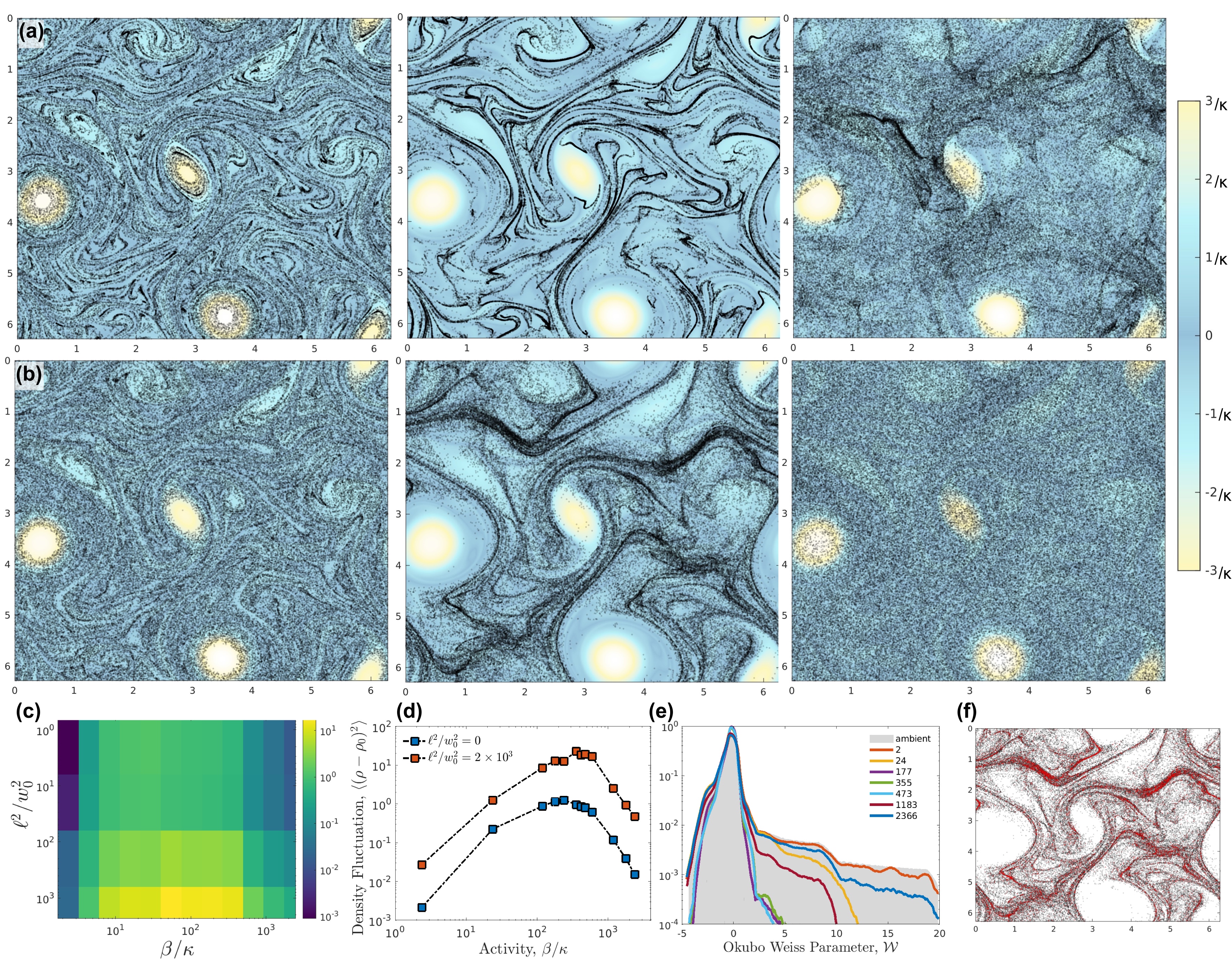}
          %\hspace{0cm} \newline 
          \caption{\label{Fig3} \textbf{Caustics in turbulent flow.} The blue-yellow colorbar represents the vorticity magnitude in the ambient turbulent flow scaled by the root-mean-square velocity gradient $\kappa = 0.8454$s$^{-1}$ 
          and black speckles are the locations of the geometric centers of $10^{5}$ active dimers with preferred length $w_{0} = 500 \mu m$. (a) $\ell^2/w_{0}^{2} = 2 \times 10^{3}$ with (from left to right) $\beta/\kappa = 23.6$,  $236$, and $2360$.
          (b) $\ell^2/w_{0}^{2} = 0$ with (left to right) $\beta/\kappa = 23.6$,  $236$, and $2360$.  
          For intermediate activity the particles display pronounced caustics [see Appendix \ref{videos} Video 3 \& Video 4].(c) heat-map of the number-density fluctuation plotted in $(\beta/\kappa, \ell^{2}/w_{0}^{2})$ plane. For comparison with plankton motility see Fig. \ref{Fig1} (a).
          (d) number-density fluctuation as a function of activity, showing pronounced agglomeration for intermediate levels of activity. (e) Okubo-Weiss parameter plotted for $\ell=0$ and various values of activity $\beta/\kappa$. (f) For $\ell^2/w_{0}^{2} = 0$, $\beta/\kappa = 236$,  the colliding pairs of particles are colored red; pronounced in caustics regions.} 
         \end{center}
    \end{figure*}
Similar to Section \ref{sbp} and \ref{hookeancloud}, we study the dynamics of a suspension of active preferred-length dimers in a point vortex [see Appendix \ref{ABPvortex}]. Setting $\ell = 0$ and $\bm{\eta =0}$, and integrating equations \eqref{eqn:dimerX5} - \eqref{eqn:dimerX6} with uniformly random initial orientations and positions, distributed on a disk of radius $4\sqrt{\tilde{\Gamma} \tau}$, gives rise to caustics in the inner region, with the motility $\beta$ playing a more conspicuous role than in the active Hookean case. Fig. \ref{fig:abp1} (a) \& (b) compare the effective centrifugation of inertial and active particles, with $\tilde{\beta}=0.5$ and $\alpha=1.0$ for various time frames, where $\tilde{\beta} \equiv \beta \sqrt{\tau/{\tilde{\Gamma}}}$ is the non-dimensional motility. The radial number density of a preferred-length dimer does not have a steady state [see Fig. \ref{fig:abp1} (c)], and particles keep drifting radially outward on average, akin to inertial particles, and unlike the Hookean dimers. A crucial distinction is the display of sharper caustics by the inertial particles, with a complete expulsion of particles within a ring that expands radially outwards with a velocity that asymptotically decays to $0$ as $r \to \infty$ \cite{Croor2015}. In contrast, the radial drift of active particles approaches a constant velocity for $r \gg 1$ [Appendix \ref{videos} Video 2]. The outer solution from the asymptotic analysis of \eqref{abp_r} - \eqref{abp_omega} [see Appendix \ref{ABPvortex}] predicts this constant radial velocity to be the non-dimensional motility $\tilde{\beta}$ of a ``free" particle independent of $\alpha$, consistent with the linearity of the rays at late times in Fig \ref{fig:abp1} (d). We also find a caustics curve in the $r-t$ plane by measuring the intersection of adjacent trajectories with $\Delta r = 0.01$, where the critical radius for caustics [defined in Fig. \ref{caustics_regime} (d)] also depends on $\alpha$ and $\tilde{\beta}$.

\section{Active Caustics in Turbulence: Direct Numerical Simulation}
To study the dynamics of a collection of ABPs in unsteady vortical flows, we write a pseudospectral code to solve the Navier-Stokes equations 
%in a  $2\pi$ periodic domain with $512^{2}$ collocation points and a 
with a deterministic external forcing $F_{0}q \cos{q x}$ on the background flow $\vect{U}$, in the stream-function-vorticity formulation \cite{gitproj}. %\remarkSR{[Would be nice to make this a bibitem but I can't figure out how.]}
In two dimensions, the inverse cascade feeds energy into long wavelengths, which we avoid by including an Ekman friction $\mu$ \cite{Boffetta2012}, in addition to the viscous dissipation $\nu$. The spectral Direct Numerical Simulations (DNS) are performed in a $2\pi$ periodic domain with $512 \times 512$ collocation points, and the parameters given in Table 1.
\begin{table}[h]
\centering
\begin{tabular}{ p{1.4cm} p{1cm} p{1.5cm} p{2.5cm} p{1.2cm} }
\hline
Domain & $k_{f}$ & $F_{0}$ & $\nu$ & $\mu$ \\
\hline
\\
\addlinespace
512$^{2}$ & 3 m$^{-1}$ & 0.1 ms$^{-2}$ & $5\times 10^{-6}$ m$^{2}$s$^{-1}$ & 0.01 s$^{-1}$ \\
\\
\hline
\end{tabular}
\caption{Spectral parameters for the Direct Numerical Simulations}\label{dns}
\end{table}
%[see Appendix \ref{DNS}]. 
This gives the flow velocity field $\vect{U}$ that drives the particle dynamics. A one-way coupling is assumed, wherein the ambient flow stirs the particles but particles do not generate flows. We use $w_{0}$ and the inverse of the root-mean-square velocity gradient 
$\kappa \equiv \sqrt{\langle \nabla \vect{U}:\nabla \vect{U}\rangle}$ 
of the background flow in the turbulent steady state as length and time scales respectively, which gives the non-dimensional parameters, $({\beta/\kappa}, \tau \kappa, \alpha, {\ell}^2/w^{2}_{0})$
and noise strength $\sqrt{2D/ w^{2}_{0}\kappa}$.
We fix $\alpha = 1$, which in the geometric interpretation of \cite{Jeffery1922} corresponds to a needle-shaped particle, and $\tau = 1$, with $\tau\kappa = 0.8454$. This leaves a two-dimensional parameter space $({\beta/\kappa}, \, {\ell}^2/w^{2}_{0})$ of activity and polar alignability, respectively.

In a turbulent steady state, we initialise the particles with uniformly random initial positions and orientations. In the steady state of particle dynamics, we find tracer-like behavior for small values of motility strength $\beta/\kappa$ in which swimmers get trapped within the vortices, consistent with the single vortex study.
For large values of $\beta/\kappa$, swimmers exhibit ballistic dynamics, leading to a homogeneous number density of particles. %The compelling features of c
Caustics appear sharply at intermediate values of $\beta/\kappa$ [see Appendix \ref{videos} Video 3], where we see preferential sampling and clustering [see Fig. \ref{Fig3} (a) \& (b)]. We quantify caustics-induced clustering by measuring the density fluctuation with respect to the initial homogeneous state [see Appendix \ref{denflu}], and find pronounced caustics 
for a range of $\ell^{2}/w_{0}^{2}$ and $\beta/\kappa$ [see Fig. \ref{Fig3} (c)]. 
We find that for intermediate values of activity $\beta/\kappa$, increasing $\ell^{2}/w_{0}^{2}$ sharpens the caustics filaments [see Appendix \ref{videos} Video 4]. The density fluctuation exhibits a peak around $\beta/\kappa \simeq 10^{2}$ [see Fig. \ref{Fig3} (d)], which is similar to the dynamics of IP, where for intermediate values of Stokes number $St$, particles display clustering \cite{EATON1994169} and sharp caustics \cite{Wilkinson_2005}, as compared to large and small values of $St$. 

\begin{figure}[t]
        \centering
        \includegraphics[width = \linewidth]{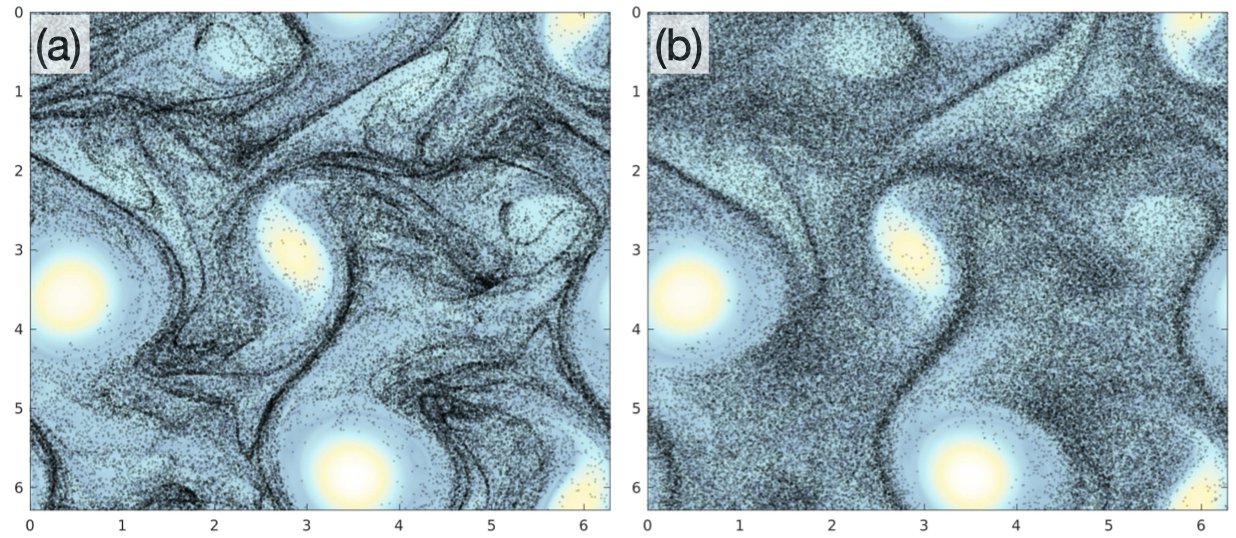}
        \caption{\textbf{Effect of Orientational Noise.} Dynamics with noise in equation \eqref{eqn:dimerX6} for (a) $\sqrt{2D/w_{0}^2\kappa} = 0.91$ and (b) $\sqrt{2D/w_{0}^2\kappa} = 9.19$, shows softening of caustics with the coarse structure retained [see Appendix \ref{videos} Video 6]. The color-bar for the background vorticity field is the same as Fig. \ref{Fig3} (a)\&(b).}\label{caustics_noise}
    \end{figure}

We quantify how swimmers sample the flow by calculating the Okubo-Weiss parameter, $\mathcal{W} = \omega^2 - 2 \bsf{S}:\bsf{S}$ at the location of particles, where $\bsf{S}$ is the symmetric part of the velocity gradient tensor \cite{Jason2018} and $\omega$ the vorticity. $\mathcal{W}>0$ implies that the particles are in a vortical region, and $\mathcal{W}<0$ indicates localization of particles in the straining region. The distribution of the Okubo-Weiss parameter $\mathcal{W}$ sampled over all particle locations gives the deviation from homogeneous sampling of the flow; when compared with the distribution of $\mathcal{W}$ over the entire flow domain [see Fig. \ref{Fig3} (e)]. We find that particles cluster preferentially in the straining regions, in a manner similar to that of gyrotactic swimmers \cite{Durham2013, PhysRevLett.116.108104}. This clustering coincides with pronounced collisions or path-crossing (see Fig. \ref{Fig3} f and Appendix \ref{videos} Video 5), that marks caustics of active particles in flow, akin to the inertial case \cite{Wilkinson_2005}, as might have been anticipated from our stationary vortex studies, and as argued to arise for gyrotactic swimmers in turbulence \cite{PhysRevLett.116.108104}. Furthermore, to demonstrate the robustness of caustics to noise in equation \eqref{eqn:dimerX6}, we include a uniformly random, uncorrelated noise $\vect{\eta}$ with zero mean and unit width. For realistic levels of non-dimensional noise strength $\sqrt{2D/w_{0}^2\kappa}$, we find softening of caustics that retains its qualitative features [see Fig. \ref{caustics_noise}, and Appendix \ref{videos} Video 6 ], and high values of noise strength $\sqrt{2D/w_{0}^2\kappa} \simeq \mathcal{O}(10^{2})$, randomizes the particle positions to produce a uniform state.

To explore the relevance of our results in marine ecosystems, we plot the typical values of $\beta/\kappa$ and $Re$ for various swimmers [see Fig. \ref{Figure1} (a)], based on published data on swimming speed and size, and typical shear-rates in the upper mixed layer of the open ocean \cite{StockerARFM2012} [See Appendix \ref{plankton}]. For small swimming organisms, like ciliates, invertebrate larvae, and copepods, when turbulence energy dissipation rates are towards the lower end of the range of values observed in the upper mixed layer of the ocean  $10^{-8} - 10^{-6}$ m$^{2}$s$^{-3}$ \cite{StockerARFM2012}, $\beta/\kappa$ can be in the intermediate regime, and from the comparison of $\beta/\kappa$ in Fig. \ref{Figure1}(a) and Fig. \ref{Fig3}(c) we expect some manifestation of active caustics in these natural biological suspensions. Interparticle collisions, which can be central to sexual reproduction in suspension of swimming organisms, formally occur in our framework through a phase-space singularity, which is missing in the prior descriptions of clustering in turbulence \cite{Durham2013, Zhan2014}.

\section{Conclusions}
We have shown how advection, rotation and alignment by an ambient vortical flow can lead to a divergence in the number density of suspended self-propelled particles without mechanical inertia. We have considered two dimer models of motile particles, Hookean and preferred-length, in vortical flows, corresponding, when noise is included to active Ornstein-Uhlenbeck \cite{Howfar2016, Bonilla2019} and active Brownian particles \cite{Romanczuk2012}. To facilitate future studies including exploration of the possible dependence of caustic structure  on the specific microswimmer model, we have made our simulation codebase publicly available in \cite{gitproj}.
% \begin{figure}[t]
%  \begin{center}
%  \includegraphics[width=8.5 cm]{mainFigures/Fig_marine_plankton.jpeg}
%  %\hspace{0cm}\newline
%  \caption{ We depict typical values of the ratio of flow time scale $1/\norm{\nabla \vect{U}}$ to swimming time scale $1/\beta$, and Reynolds number $Re$ for a marine bacterium \textit{Vibrio alginolyticus} \cite{%10.7554/eLife.22140, 
%  doi:10.1073/pnas.1602307113}, various dinoflagellates, ciliates \cite{Lauga2019}, invertebrate larvae \cite{FUCHS2016109} and copepods \cite{FUCHS2016109}. For $\norm{\nabla \vect{U}}$ we substitute the Kolmogorov shear rate for the range of previously measured energy dissipation rates in the upper mixed layer of the ocean $10^{-8} - 10^{-6}$ m$^{2}$s$^{-3}$ \cite{StockerARFM2012} [see Appendix \ref{plankton}].}\label{natural_swimmers} %\remarkSR{Check source for coffee photo or take a photo yourself.}}  -- This is my own coffee mug :) 
%  \label{Figure6}
%  \end{center}
%  \end{figure}

By recasting the dynamics of motile inertialess particles in a form close to that of the classic Maxey-Riley equations of inertial particles (IPs), we highlight the similarities and differences between the centrifugation of these two types of particle, in the illustrative setting of a single point vortex. We show the formation of caustics in both the dimer types mentioned above, by analysing the intersection of rays in the $r-t$ plane. For a range of the flow-coupling parameter $\alpha$, we demarcate the regimes in the ${\alpha}-r$ plane where caustics occur. We study the effect of advection by more general vortical flows in the form of two-dimensional Navier-Stokes turbulence generated by direct numerical simulation, characterizing the preferential sampling of straining regions by the swimmers by means of the Okubo-Weiss parameter. We find that for intermediate values of the dimensionless motility $\beta/\kappa$, clustering and caustics are more pronounced, similar to the dynamics of IP as the Stokes number $\text{St}$ is varied, suggesting that $\beta/\kappa$ plays the same role as $\text{St}$. This hitherto unexplored caustics regime is of interest formally because crossing of active-particle worldlines renders their velocity field multiple-valued, and practically because the resulting rapid, transient collisions offer a strikingly effective natural mechanism for close encounters between organisms. These encounters can increase mating opportunities or contact-dependent interactions at low mean concentrations, beyond the effects of elevated local concentration \cite{Durham2013, Zhan2014}. The eco-physiological impact of clustering and caustics is expected to be organism-specific and awaits experimental verification.

%\rc{beyond what is predicted by generic clustering \cite{Durham2013, Zhan2014}, which only elevates local concentration. The precise eco-physiological impact of clustering and caustics is expected to be organism-specific and remains to be tested experimentally.}

%\remarkSR{[As we discussed earlier, others have already argued that flow-induced concentration achieves this. Are our caustics really that much better?]} \remarkRC{1) previous literature do not show singular concentration, 2) generic clustering can bring particles close, but not make them bump into one another rapidly. Both of these can independently have implications in the marine biological world.}

The biological implications of active caustics in marine ecosystems may take time to be fully understood, but the theoretical framework developed here can be of immediate value for conceptualizing field observations. As mentioned in the beginning of this article, in complex natural environments, neither inertia nor activity can be directly controlled, and collisional aggregation \cite{pumir2016} can arise from both inertial and active caustics. Disentangling their respective contributions, both their similarities and distinctions, is therefore essential for interpreting swimmer dynamics in field settings. A unified mechanistic framework for particle aggregation (and disaggregation) in marine environments \cite{Burd2009} remains an open problem, 
to which our article contributes new insights. Our study reveals the minimal conditions for the occurrence of active caustics: (i) activity acts as a singular perturbation to the dynamics, that is, an arbitrarily small level of activity is sufficient to generate caustics, (ii) vortical flows with Jeffery-type coupling between swimmer orientation and flow ($\alpha \neq 0$) are adequate to induce caustics; turbulence is not a prerequisite, and (iii) active caustics can arise even in the absence of organism-specific behaviors such as gyrotaxis or chemotaxis, which are not included in our model. We further demonstrate that (iv) fore–aft asymmetry (characterized by $\ell \neq 0$) and particle rigidity are not essential, and that (v) caustics remain robust to orientational noise. In three dimensions, accumulation is known to be weaker than in two dimensions \cite{Zhan2014}, but we expect the criteria for caustics formation to remain valid, with extremal surfaces rather than curves. While optical caustics can be engineered \cite{Zhou2024}, whether an analogous framework can be developed for controlling active suspensions remains a question.

Although coarse-graining eliminates multivaluedness of the velocity field, the accompanying singularity in the density field persists. In natural systems, however, the divergence in particle number density would likely be regularised by inter-particle hydrodynamic, steric and/or behavioral interactions. Whether condensation driven by caustics in active suspensions constitute a bona fide nonequilibrium phase transition, and whether active turbulence generated by swimming microbes can produce caustics in their own concentration fields, remains to be explored.

\medskip
%\noindent \textbf{METHODS}
\section{Acknowledgements}
SR acknowledges a JC Bose Fellowship from the ANRF, India, support from the Tata Education and Development Trust, an ICTS Endowed Visiting Professorship, and discussions in the Program on Complex Lagrangian Problems of Particles in Flows, ICTS-TIFR, Bangalore, and RC \& RG acknowledge support from the Department of Atomic Energy, Government of India, under project no. RTI4001. SR thanks the Isaac Newton Institute for Mathematical Sciences, Cambridge, for support and hospitality during the programme Anti-diffusive dynamics: from sub-cellular to astrophysical scales where work on this paper was undertaken, supported by EPSRC grant no EP/R014604/1. R thanks Saumav Kapoor and Samriddhi Sankar Ray at ICTS for insightful discussions. RC acknowledges support from the International Human Frontier Science Program Organization, and discussions in Active Matter in Complex Environments Conference at the Aspen Center for Physics, and thanks Prakash Lab for valuable insights into marine ecology.

\medskip
\begin{appendices}
\appendix

\section{Supplementary Videos}
\label{videos}
\textbf{Video 1:} Active Hookean dimers with $\alpha =1$, $\lambda = 0$ in a point vortex flow. It is compared with the dynamics of inertial particles.

\textbf{Video 2:} Active Preferred-length dimers with $\alpha =1$, $\lambda = 0$, $\beta \sqrt{\tau/\Gamma} = 0.5$ in a point vortex flow.

\textbf{Video 3:} Preferred-length dimers in turbulence with flow parameters given by Table I of methods section, and with $\alpha = 1$, $\tau = 1$,  $\ell^{2}/w_{0}^{2} = 0$, $\beta/\kappa = 23.6$, $236$, and $236$ (increasing from left to right). The intermediate values of activity $\beta/\kappa$ presents the regime of most pronounced caustics as shown in Fig. 3 (b). 

\textbf{Video 4:} Preferred-length dimers in turbulence with $\alpha = 1$, $\tau = 1$, $\beta/\kappa = 236$, and $\ell^{2}/w_{0}^{2} = 0$ (left) and  $\ell^{2}/w_{0}^{2} = 2 \times 10^{3}$ (right). Increasing the polar aligning parameter $\ell^{2}/w_{0}^{2}$ intensifies the caustics [comparison between the middle column of Fig. 3 (a) \& (b)].

\begin{table*}[t]
 \begin{center}
 \label{table_organism}
\begin{tabular}{ |p{2.5cm}| p{2cm}|p{2cm}|p{2cm}| p{2.5cm} | p{5cm} |  }
 \hline
 \multicolumn{6}{|c|}{\textbf{Data used to make Fig.1 (a)}} \\
 \hline
 Creature& size ($\mu$m) & Speed ($\mu$m s$^{-1}$) & $\beta$ & Re & Data Source\\
 \hline
  Dinoflagellate & 63.7 & 261.6 & 9.7 & 0.01 & \href{https://doi.org/10.7554/elife.44907}{M. Lisicki et al., eLife 8:e44907 (2019).}\\
  Ciliate & 180.2 & 1184.2 & 13.1 & 0.2 &  \href{https://doi.org/10.7554/elife.44907}{M. Lisicki et al., eLife 8:e44907 (2019).}\\
  Larvae & 396.6 & 2275 & 8.2 & 0.9 & 1) \href{https://doi.org/10.1016/j.pocean.2015.12.010}{H. L. Fuchs and G. P. Gerbi, Progress in Oceanography,
vol. 141, pp. 109–129, 2016}, 2) \href{https://doi.org/10.2307/1542690}{D. Wendt, The
Biological Bulletin, vol. 198, no. 3, pp. 346–356, 2000}, 3) \href{https://doi.org/10.4319/lo.2004.49.6.1937}{ H. L. Fuchs et al., Limnology and Oceanography, vol. 49, no. 6,
pp. 1937–1948, 2004} \\
  Copepod & 1284 & 3440 &2.5 & 4.4 & \href{https://doi.org/10.1016/j.pocean.2015.12.010}{H. L. Fuchs and G. P. Gerbi, Progress in Oceanography,
vol. 141, pp. 109–129, 2016}\\
  Marine Bacteria & 3 & 40	& 13.3	& 0.0001 & 1) \href{https://doi.org/10.7554/eLife.22140}{ M. Chen et al. eLife, vol. 6, p. e22140, jan 2017.} 2) \href{https://doi.org/10.1073/pnas.1602307113}{ K. Son et al.
PNAS, vol. 113, no. 31, pp. 8624–8629, 2016} \\
 \hline
\end{tabular}
\end{center}
\end{table*}

\textbf{Video 5:} Preferred-length dimers in turbulence with $\alpha = 1$, $\tau = 1$, $\beta/\kappa = 236$, and $\ell^{2}/w_{0}^{2} = 0$, where the pairs of particles whose trajectories intersect within the numerical time step $10^{-3}$ s are coloured red; showing extreme path-crossing events in the regions of high number density.

\textbf{Video 6:} Preferred-length dimers in turbulence with $\alpha = 1$, $\tau = 1$, $\beta/\kappa = 236$, $\ell^{2}/w_{0}^{2} = 0$,  and non-dimensional noise strength $\sqrt{2D/w_{0}^2\kappa} = 9.19$.

\section{Plotting Reynolds number and \texorpdfstring{$\beta/\norm{\nabla \vect{U}}$}{Lg} for various marine organisms}
\label{plankton}
We use the kinematic viscosity of water, $\nu = 10^{-6}$ m$^{2}$s$^{-1}$ with previously measured energy dissipation rates in the upper mixed layer of the ocean, $\epsilon = 10^{-8} - 10^{-6}$ m$^{2}$s$^{-3}$ \cite{StockerARFM2012}, to calculate the Reynolds number and shear-rate using the relation $\langle \norm{\nabla \vect{U}}\rangle_{rms} = (\epsilon/\nu)^{1/2}$.

The Table \ref{table_organism} is a compilation of data-sets published previously by other researchers (source is given in the table). Each creature type exhibits a distribution of size and swimming speed. We use the average value from the known data. There is limited data on the size and swimming statistics of marine bacteira; we use the data for \textit{Vibrio alginolyticus}, which is studied due to its bio-medical importance.

\section{Active Hookean Dimer in a Point Vortex}
\label{AOUPvortex}
In the absence of external force field $\vect{F}$, and redefining a flow-dependent relaxation time, $\tilde{\tau}^{-1} \equiv \bsf{I}/\tau - ( \alpha \, \bsf{S} + \bsf{A} \,)$, gives the equation
\begin{equation}
\frac{d \mathbf v}{dt} - \mathbf{v}\cdot\nabla \mathbf{U} = \,(\mathbf{U} - \mathbf{v}) \cdot \tilde{\tau}^{-1}.
\label{eqn:4}
\end{equation}
The velocity field generated by a point-vortex in polar coordinates is 
\begin{equation}
\mathbf{U} = \frac{\Gamma}{2\pi r} \hat{\theta} \equiv \frac{\bar{\Gamma}}{r} \bf{\hat{\theta}}.
\label{eqn:5}
\end{equation}

% \begin{figure}[t]
%       \begin{center}
%       \includegraphics[width=8.5 cm]{Supplementary_figures/ABP1.jpg}
%       %\hspace{0cm}\newline
%       \caption{\label{fig:abp1}\textbf{Centrifugation and caustics of persistent active dimers (noiseless ABP) in a point vortex}: (a) \& (b) show time frames of particle positions around a point vortex for inertial particles and noiseless ABP respectively, shown for $\tilde{\beta} = 0.5$ and $\alpha=1.0$ in \eqref{eqn:dimerX5s} - \eqref{eqn:dimerX6s}. (c) radial number-density of active particles (green) compared with inertial particles (purple) shown for time $10\tau$, and the number-density at $t=0$ (grey). Arrows schematically depict the radial drift of the unsteady state. (d) Intersections of adjacent trajectories marking the caustics curve (red circles) obtained for $\tilde{\beta} = 0.5$, $\alpha = 1.0$ and $\Delta r = 0.01$.} 
%       %(d) Caustics phase diagram in the $R_0-\alpha$ plane.}
%       \end{center}
% \end{figure} 

In polar coordinates the position derivatives are
\begin{align}
 \mathbf{X} & = r \mathbf{\hat{r}}, \nonumber \\ 
\frac{d \mathbf{X}}{dt} & = \frac{dr}{dt} \mathbf{\hat{r}} \, + \, r\frac{d\theta}{dt} \bf{\hat{\theta}}, \label{eqn:6a}\\
{\rm and} \quad \frac{d^{2} \mathbf{X} }{dt^{2}} &= \frac{d^{2} r}{dt^{2}} \mathbf{\hat{r}} + 2 \frac{d r}{dt}\frac{d \theta}{dt} \mathbf{\hat{\theta}} - r \left(\frac{d \theta}{dt}\right)^{2} \mathbf{\hat{r}} + r\frac{d^{2} \theta}{dt^{2}} \mathbf{\hat{\theta}}. \label{eqn:7a}
\end{align}

To demonstrate the emergence of an effective centrifugal force due to the coupling of activity with the background flow, we neglect the external force, and the Gaussian white noise in \eqref{eqn:inertialEoM}, and consider fore-aft symmetric particles such that $\ell = 0$. For a point vortex, the antisymmetric tensor $\bsf{A}$ is zero everywhere except at the origin, and the symmetric part is
\begin{align}
\bsf{S} = \left(\begin{array}{cc} 0 & -\tilde{\Gamma}/r^{2}\\ -\tilde{\Gamma}/r^{2} & 0 \end{array}\right).
\label{eqn:8}
\end{align}
Using \eqref{eqn:5}-\eqref{eqn:8} in \eqref{eqn:4} gives 
\begin{align}
\tau \left[\frac{d^{2} r}{dt^{2}} \mathbf{\hat{r}} + 2 \frac{d r}{dt}\frac{d \theta}{dt} \mathbf{\hat{\theta}} - r \left(\frac{d \theta}{dt}\right)^{2} \mathbf{\hat{r}} + r\frac{d^{2} \theta}{dt^{2}} \mathbf{\hat{\theta}} + \frac{\tilde{\Gamma}}{r} \frac{d \theta}{dt} \mathbf{\hat{r}} + \frac{\tilde{\Gamma}}{r^2} \frac{d r}{dt} \mathbf{\hat{\theta}} \right] & \nonumber \\
= \frac{\bar{\Gamma}}{r} \mathbf{\hat{\theta}} - \frac{dr}{dt} \mathbf{\hat{r}} \, - \, r\frac{d\theta}{dt} \mathbf{\hat{\theta}} 
  +  \frac{\alpha \tau \tilde{\Gamma}}{r^{3}}  \mathbf{\hat{r}} - \frac{\tau \alpha \tilde{\Gamma}}{r} \frac{d\theta}{dt} \mathbf{\hat{r}} - \frac{\tau \alpha \tilde{\Gamma}}{r^{2}} \frac{dr}{dt} \mathbf{\hat{\theta}}.
\label{eqn79}
\end{align}
Separating equations in the $\mathbf{\hat{r}}$ and $\mathbf{\hat{\theta}}$ directions in \eqref{eqn79}, we get two coupled equations
\begin{equation}
\label{eom}
\tau \frac{d^{2} r}{dt^{2}} - \tau r \left(\frac{d \theta}{dt}\right)^{2} = - \frac{dr}{dt} + \frac{\alpha \tau \tilde{\Gamma}^{2}}{r^{3}} - \frac{\tau (1+\alpha)\tilde{\Gamma}}{r} \frac{d \theta}{dt},
\end{equation}
\begin{equation}
2 \tau \frac{d r}{dt}\frac{d \theta}{dt} + \tau r\frac{d^{2} \theta}{dt^{2}} = \frac{\bar{\Gamma}}{r} - r\frac{d\theta}{dt} - \frac{\tau (1+\alpha) \tilde{\Gamma}}{r^{2}} \frac{dr}{dt}.
\end{equation}
% \rishi{Maybe we should non-dimensionalize the equations earlier? }
Choosing $\sqrt{\tilde{\Gamma}\tau}$ and $\tau$ as the length and time scales respectively, we arrive at the following non-dimensional equations involving non-dimensional variables $r,t$ and $L = r^2 \dot{\theta}$.
\begin{align}
\label{nd_eom}
    \begin{split}
    \ddot{r} - \frac{L^2}{r^3} &= -\dot{r} + \frac{\alpha - (1 +\alpha)L }{r^3}\\
    \dot{L} &= 1- L - (1+\alpha) \frac{\dot{r}}{r} \ ,
    \end{split}
\end{align}
where $\dot{(.)}\equiv d(.)/dt$.

\section{Active Preferred-length Dimer in a Point Vortex}
\label{ABPvortex}
The equations for an active dimer which has a preferred length are given by
\begin{align}
\dot{\mathbf{X}} & =  \mathbf{U} + \beta \mathbf{w}% \equiv \vect{v},   
\label{eqn:dimerX5s}\\
\dot{\mathbf{w}} & = \frac{1}{\tau}\left(1 - \frac{|\mathbf{w}|^2}{w_{0}^{2}} \right)   \mathbf{w} + (\alpha \bsf{S} + \bsf{A})\cdot\mathbf{w} \nonumber \\ & - {\ell}^{2} \nabla^{2} \vect{U} \, %+ {\sqrt{2D} \over \tau}\bm{ \eta}(t)\label{eqn:dimerX6}.
+ \sqrt{2D}\bm{ \eta}(t). \label{eqn:dimerX6s}%{+ {\bf f}(t)},\label{eqn:dimerX6s}
\end{align}
With $\ell = 0$ and $\bm{ \eta} =0$,  \eqref{eqn:dimerX5s} and \eqref{eqn:dimerX6s} can be recast as
\begin{equation}
\frac{d \mathbf v}{dt} - \mathbf{v}\cdot\nabla \mathbf{U} = \,(\mathbf{U} - \mathbf{v}) \cdot \tilde{\tau}^{-1}
\label{eqn:36}
\end{equation}
where $\tilde{\tau}^{-1} \equiv \frac{\bsf{I}}{\tau} \left(\frac{||V-U||}{\beta^{2}} - 1 \right) - ( \alpha \, \bsf{S} + \bsf{A} \,)$. The non-dimensional equations with $R = \sqrt{\tilde{\Gamma} \tau}$ and $T = \tau$, in cylindrical polar coordinates become
\begin{align}
\tilde{\beta}^2 \frac{d^{2} r}{dt^{2}} & = \tilde{\beta}^2 r \left( \frac{d\theta}{dt}\right)^{2} \nonumber \\
& + \frac{dr}{dt} \left\{ 2\frac{d\theta}{dt} -  \left( \frac{dr}{dt}\right)^{2} -r^{2} \left( \frac{d\theta}{dt}\right)^{2} + \tilde{\beta}^2 - \frac{1}{r^{2}} \right\} \nonumber \\
& + \frac{\tilde{\beta}^2 \alpha}{ r^{3}}  - \frac{\tilde{\beta}^2 (1+\alpha)}{ r} \frac{d\theta}{dt},
\end{align}
\begin{align}
\tilde{\beta}^2 r \frac{d^{2} \theta}{dt^{2}} & = - 2 \tilde{\beta}^2 \frac{d\theta}{dt}\frac{d\theta}{dt}  + + \left( r \frac{d\theta}{dt} - \frac{1}{r} \right) \times \nonumber \\
& \left\{ 2\frac{d\theta}{dt} -  \left( \frac{dr}{dt}\right)^{2} -r^{2} \left( \frac{d\theta}{dt}\right)^{2} + \tilde{\beta}^2 - \frac{1}{r^{2}} \right\} \nonumber \\
& - \frac{\tilde{\beta}^2 (1+\alpha)}{ r^{2}} \frac{dr}{dt}, 
\end{align}

\begin{figure}[b]
\begin{center}
\includegraphics[width=8.5 cm]{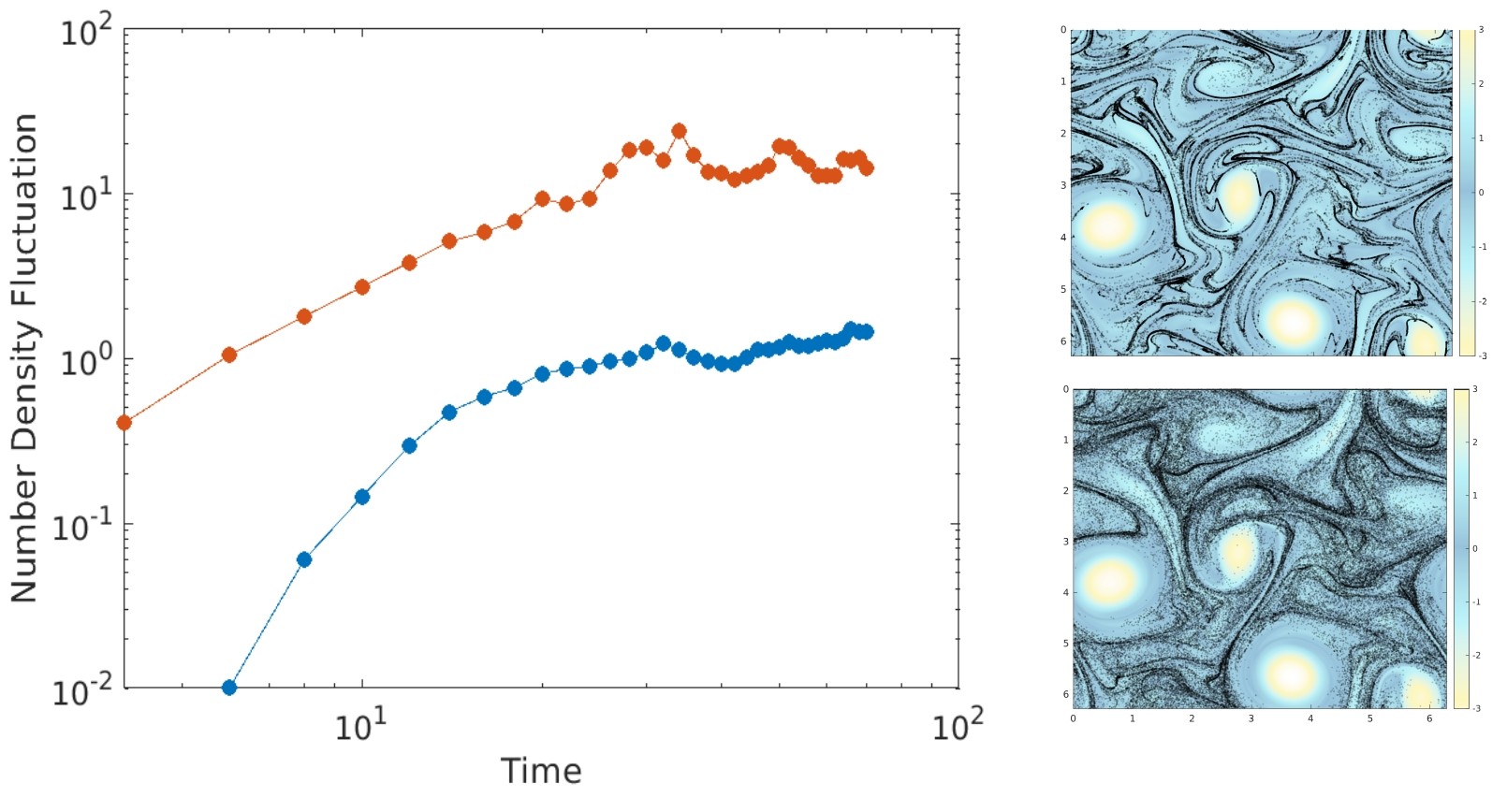}
%\hspace{0cm}\newline
\caption{\label{fig:3}\textbf{Number-density fluctuations:} The fluctuations saturate as a function of time as the particulate structure evolves from a uniformly random state to exhibiting caustics. Red and blue curves correspond to $\ell^2/w_{0}^{2} = 2 \times 10^{3}$ and $\ell^{2}/w_{0}^{2} = 0$ respectively. The activity $\beta/\kappa = 236$ in both cases. The top and bottom panels on the right correspond to the top and bottom curves on the left respectively. }
\end{center}
\end{figure}

\begin{figure}[b]
\begin{center}
\includegraphics[width=8.5cm]{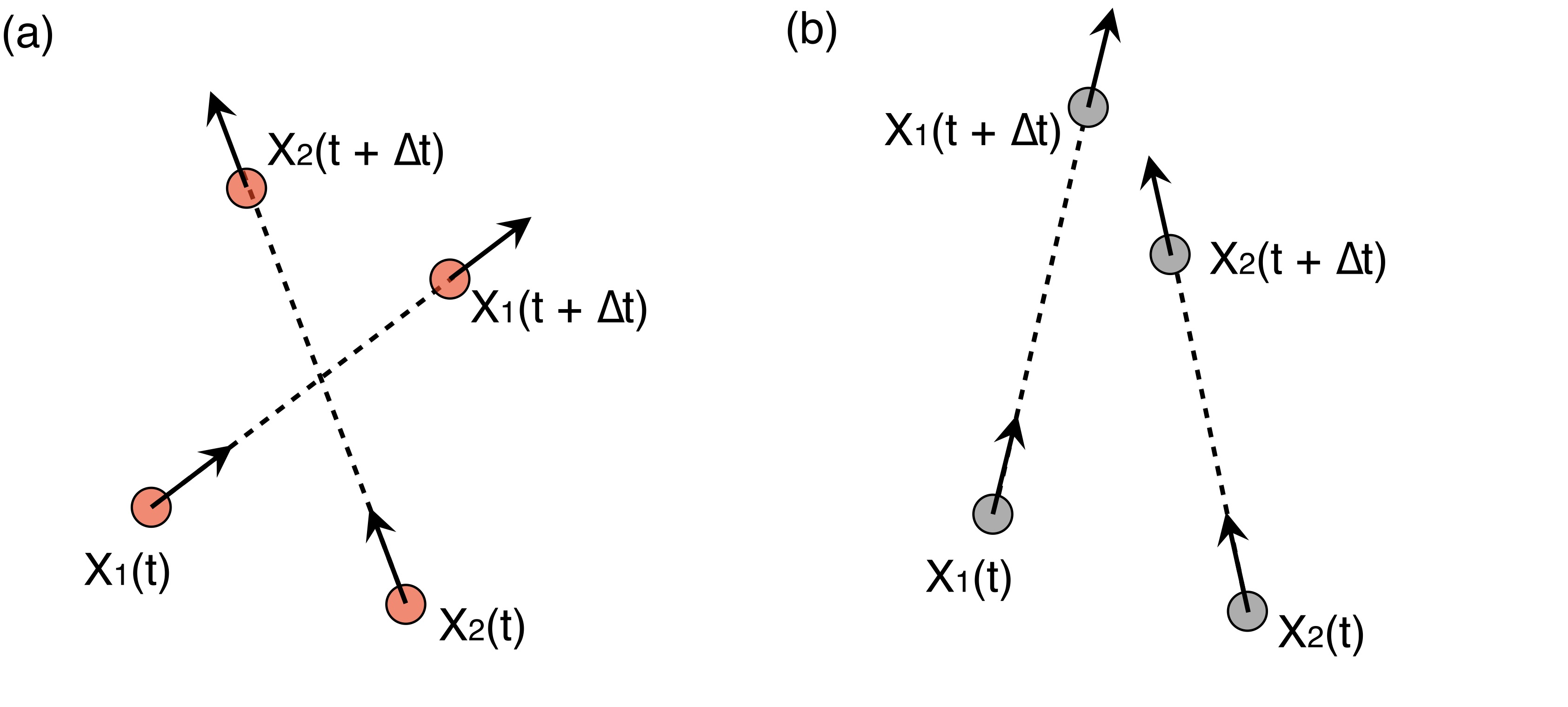}
%\hspace{0cm}\newline
\caption{\label{fig:collision}\textbf{Collisions:} A schematic of (a) colliding (red) and (b) non-colliding (grey) pairs of particles, where the intersection of paths within the numerical time step $\Delta t$ marks the collision event.}
\end{center}
\end{figure}

where $\tilde{\beta} \equiv \beta \sqrt{\tau/{\tilde{\Gamma}}}$ is the non-dimensional motility. These may be recast as a system of first-order equations as
\begin{align}
%\label{total_sol}
\frac{dr}{dt} & = v_{r}, \label{abp_r}\\
\tilde{\beta}^2 \frac{dv_{r}}{dt} &= \tilde{\beta}^2 r\omega^{2}  + v_{r} \left\{ 2\omega -  {v_{r}}^{2} -r^{2} {\omega}^{2} + \tilde{\beta}^2 - \frac{1}{r^{2}} \right\} \nonumber \\
& +  \frac{\tilde{\beta}^2 \alpha}{r^{3}} - \frac{\tilde{\beta}^2 (1+\alpha)\,\omega}{r}, \label{abp_v}\\
\frac{d\theta}{dt} & = \omega, \label{abp_theta}\\
\tilde{\beta}^2\frac{d\omega}{dt} & = - \frac{2 \tilde{\beta}^2\, v_{r} \, \omega}{r}  \nonumber \\
& + \left( \omega - \frac{1}{r^{2}} \right) \left\{ 2\omega -  {v_{r}}^{2} -r^{2} {\omega}^{2} + \tilde{\beta}^2 - \frac{1}{r^{2}} \right\} \nonumber \\
& - \frac{\tilde{\beta}^2(1+\alpha)\, v_{r}}{r^{3}} \label{abp_omega}.
\end{align}

By calculating the divergence of crossing time, for a fixed $\tilde{\beta} =1.0$, we get the caustics phase in the $r-\alpha$ plane, similar to the hookean case. We find that rigid spheroidal microswimmers  \cite{Stark2012, Stark2016} in vortical flow also display caustics, which we do not explore here.

\section{Number density fluctuation, clustering and collisions}
\label{denflu}
Local number density fluctuations can be used as a statistical measure for the intensity of caustic-induced clustering. The space is numerically discretized into $N \times N$ cells such that there is on an average 1 particle per cell in the initial uniformaly random state
\begin{equation}
    {\Delta \rho_{ij}}^{2} = \frac{1}{N^2}\sum_{i,j} (\rho_{ij} - 1)^{2}.
\end{equation}
Here, $i,j$ are the cell index corresponding to a given spatial location. The initial state itself contributes a residual density fluctuation, which we subtract to obtain purely caustic-induced clustering. 
To measure collisions, we find pairs of particles whose trajectories intersect within the numerical time step $\Delta t = 10^{-3}$, as shown in \ref{fig:collision}. Such particles are colored red in Supplementary Video 5; where we one finds that the regions with high number-density coincide with the crossing of particle trajectories.

\end{appendices}

%\LaTeX{} \cite{latex2e} is a set of macros built atop \TeX{} \cite{texbook}.
\bibliographystyle{apsrev4-2}
\bibliography{active_caustics}

\end{document}